# Three-dimensional spectral analysis of compositional heterogeneity at Arruntia crater on (4) Vesta using Dawn FC

Guneshwar Thangjam[1,2], Andreas Nathues[1], Kurt Mengel[2], Michael Schäfer[1], Martin Hoffmann[1], Edward A. Cloutis[3], Paul Mann[3], Christian Müller[2], Thomas Platz[1], Tanja Schäfer[1]

[1]Max-Planck-Institute for Solar System Research, Justus-von-Liebig-Weg 3, 37077, Göttingen, Germany

[2]Clausthal University of Technology, Adolph-Roemer-Straße 2a, 38678, Clausthal-Zellerfeld, Germany

[3]University of Winnipeg, 515 Portage Avenue Winnipeg, Manitoba, Canada R3B 2E9

**Editorial correspondence to:**
Guneshwar Thangjam
Justus-von-Liebig-Weg 3
37077 Göttingen
Germany.
Email: thangjam@mps.mpg.de

**Abstract**

We introduce an innovative three-dimensional spectral approach (three band parameter space with polyhedrons) that can be used for both qualitative and quantitative analyses improving the characterization of surface heterogeneity of (4) Vesta. It is an advanced and more robust methodology compared to the standard two-dimensional spectral approach (two band parameter space). The Dawn Framing Camera (FC) color data obtained during High Altitude Mapping Orbit (resolution ~ 60 m/pixel) is used. The main focus is on the howardite-eucrite-diogenite (HED) lithologies containing carbonaceous chondritic material, olivine, and impact-melt. The archived spectra of HEDs and their mixtures, from RELAB, HOSERLab and USGS databases as well as our laboratory-measured spectra are used for this study. Three-dimensional convex polyhedrons are defined using computed band parameter values of laboratory spectra. Polyhedrons based on the parameters of Band Tilt ($R_{0.92\mu m}/R_{0.96\mu m}$), Mid Ratio (($R_{0.75\mu m}/R_{0.83\mu m}$)/($R_{0.83\mu m}/R_{0.92\mu m}$)) and reflectance at 0.55 µm ($R_{0.55\mu m}$) are chosen for the present analysis. An algorithm in IDL programming language is employed to assign FC data points to the respective polyhedrons. The Arruntia region in the northern hemisphere of Vesta is selected for a case study because of its geological and mineralogical importance. We observe that this region is eucrite-dominated howarditic in composition. The extent of olivine-rich exposures within an area of 2.5 crater radii is ~ 12% larger than the previous finding (Thangjam et al., 2014). Lithologies of nearly pure CM2-chondrite, olivine, glass, and diogenite are not found in this region. Although there are no unambiguous spectral features of impact melt, the investigation of morphological features using FC clear filter data from Low Altitude Mapping Orbit (resolution ~ 18 m/pixel) suggests potential impact-melt features inside and outside of the crater. Our spectral approach can be extended to the entire Vestan surface to study the heterogeneous surface composition and its geology.

**1.0 Introduction**

Vesta is one of the most geologically interesting small bodies in the Main Asteroid Belt that bears characteristics similar to the terrestrial planets (e.g., Keil, 2002; Jaumann et al., 2012; Russell et al., 2013). Vesta is an intact differentiated object that most probably survived catastrophic impact events in the early solar system (e.g., Russell et al., 2012, 2013). The geologic study of such a body allows us to understand the evolution of planetary bodies with a silicate-dominated shell. The knowledge about Vesta has been improved by an integrated study of HED meteorites (e.g., Mittlefehldt et al., 1998; Mittlefehldt, 2015; McSween et al., 2011) and ground-based telescopic observations (e.g., Thomas et al., 1997; Hicks et al., 2014; Hiroi et al., 1995). The investigations using Dawn spacecraft images mark a significant progress because of its higher spatial and spectral resolution data compared to all former observations. The spacecraft spent more than a year in orbit around Vesta (July 2011 - September 2012) and acquired images from various distances (e.g., Russell et. al., 2012, 2013). The Framing Camera (FC) and the Visible and Near-infrared Spectrometer (VIR) are imaging instruments onboard Dawn (e.g., Sierks et al., 2011; De Sanctis et al., 2011). The Framing Camera houses seven color filters in the wavelength range from 0.4 to 1.0 μm, and a clear filter. The spatial resolution of FC color data is about 2.3 times higher than the resolution of VIR cubes (Sierks et al., 2011). The FC color data is used in this study.

**1.1 Composition and geology: pre- and post-Dawn**

Before the arrival of Dawn at Vesta, many researchers have studied this object using Earth-based telescopes. Non-uniform reflectivity and albedo variations across the surface have long been recognized (e.g., Bobrovnikoff, 1929; Gehrels, 1967). McCord et al. (1970) observed the spectral similarity of Vesta with the eucrite Nuevo Laredo in the visible and near-infrared region (0.3-1.1 μm). Later on, the spectral similarity between eucrite meteorites and Vesta were confirmed (e.g., Feierberg and Drake, 1980). Since then, the evidence for a Vestan origin of the HED suite of meteorites has increased substantially. This hypothesis was supported by the finding of Vestoids (V-type asteroids) that are spectrally similar to Vesta and located in the 3:1 orbital resonance region (e.g., Binzel and Xu, 1993; Thomas et al., 1997). A crude geochemical map (Gaffey, 1983) and a pyroxene-distribution map (Dumas et al., 1996) indicated albedo and lithological variations. A schematic map of major lithologic units of diogenites and eucrites including an olivine-bearing unit was produced by Gaffey (1997) based on rotationally-resolved ground-based telescopic data. He noted hemispheric and sub-hemispheric compositional variations. A geologic map of Vesta was constructed by Binzel et al. (1997) using Hubble Space Telescope data/Wide Field Planetary Camera (HST/WFPC2). These observations suggested that the eastern hemisphere is spectrally more diverse than the western hemisphere, dominated by units similar to diogenites and eucrites, respectively. Li et al. (2010) presented albedo and color ratio maps of the southern hemisphere using HST/WFPC2 images. They suggested that the brighter eastern hemisphere has a deeper 1-µm absorption band than the darker western hemisphere. Reddy et al. (2010) investigated the southern hemisphere of Vesta using rotationally resolved ground-based telescopic data. These authors noted that the composition in the southern hemisphere ranges from diogenites to cumulate eucrites whereas the northern hemisphere is dominated by cumulate to basaltic eucrites. Olivine was initially not detected (Reddy et al., 2010; Li et al., 2010), despite the predictions of its presence by previous investigators (Gaffey, 1997; Binzel et al., 1997; McFadden et al., 1977).

Analyzes of higher resolution Dawn images have led to a remarkable progress. A general compositional and lithological heterogeneity are mapped in terms of HED lithologies (Reddy et al., 2012b; De Sanctis et al., 2012a; Ammannito et al., 2013b; Thangjam et al., 2013; Nathues et al., 2014, 2015; Schaefer et al., 2014). The surface of Vesta is found to be dominantly eucrite-rich howardite (e.g., Thangjam et al., 2013; Ammannito et al., 2013b). Diogenite-rich lithologies are spotted inside the Rheasilvia basin and in a few localized regions outside the basin, while the equatorial regions are rather eucritic in composition (e.g., Reddy et al., 2012b; De Sanctis et al., 2012a; Ammannito et al., 2013b; Thangjam et al., 2013). Pieters et al. (2012) suggest that space weathering differs from the lunar style and is minimal compared to other airless planetary bodies. The heterogeneous composition of Vesta includes several additional components mixed in with the HED lithology, i.e., dark material (McCord et al., 2012; Reddy et al., 2012c; Palomba et al., 2014; Nathues et al., 2014), bright material (Longobardo et al., 2014; Zambon et al., 2014), glass/shock/impact-melts or orange material (Le Corre et al., 2013; Schaefer et al., 2014; Ruesch et al., 2014b) and olivine-rich material (Ammannito et al., 2013a; Thangjam et al., 2014; Ruesch et al., 2014b; Poulet et al., 2014; Nathues et al., 2015; Palomba et al., 2015).

Dark material (DM) affects Vesta's surface reflectivity significantly (e.g., McCord et al., 2012; Reddy et al., 2012c; Jaumann et al., 2012). Dark material is likely exogenic in origin, and its spectral parameters resemble carbonaceous chondrites, in particular the CM2-chondrites (McCord et al., 2012; Reddy et al., 2012; Nathues et al., 2014; Palomba et al., 2014). The CM2-chondrites along with other chondritic materials are found in variable proportions as xenoliths in HED meteorites, for instance the PRA 04401 howardite contains about 40-50% CM2-material (e.g., Herrin et al., 2011; Wee et al., 2010). Detection of OH-bearing minerals (De Sanctis et al., 2012b) in particular serpentine (Nathues et al., 2014) and elevated H-abundance (Prettyman et al., 2012) in DM regions further supports an exogenic source.

Bright material (BM) showing unusual high reflectivity is frequently encountered on Vesta (Zambon et al., 2014). Zambon et al. (2014) have analyzed sites of BM using VIR data and found that this material is spectrally consistent with pyroxenes in HEDs showing deeper band depths. Regions with BM are interpreted to be generally fresh which are neither contaminated by howarditic material through continuous impact gardening nor by exogenic carbonaceous chondritic material (Zambon et al., 2014; Reddy et al., 2012b).

Olivine-rich sites are identified for the first time at Arruntia and Bellicia craters in Vesta's northern latitudes using VIR data (Ammannito et al., 2013a). Further, such exposures are found in far northern regions (e.g., Pomponia crater) and in a few spots in the equatorial region, including potential sites at the Rheasilvia basin rim/wall (Matronalia Rupes), on crater floors (Severina crater, Tarpeia crater) and on crater central peaks (Nathues et al., 2015; Ruesch et al., 2014b; Palomba et al., 2015). Olivine was not detected initially during the early stages of Dawn observations despite the predictions of significant olivine exposures in the south polar basin (e.g., Thomas et al., 1997; Gaffey, 1997; Beck and McSween 2010; Jutzi and Asphaug, 2011; Jutzi et al., 2013; Ivanov and Melosh, 2013; Tkalcec et al., 2013), and therefore, concerns raised about the missing mantle material of Vesta (e.g., McSween et al., 2014). Based on linear spectral unmixing analysis, Combe et al. (2015) argue that a mixture of low- and high-Ca pyroxenes (hypersthene, pigeonite, and diopside) is an alternative to the presence of olivine in these regions.

An integrated analysis of Dawn FC, VIR and GRaND data, Le Corre et al. (2013) find that 'Orange Material' (OM) that appeared orange in the 'Clementine' RGB ratio image is analogue to material of impact-melt/shock/glass on Vesta's surface. They suggest that a steeper visible slope (red visible slope) is a diagnostic spectral parameter for this material. This parameter is referred elsewhere to detect and map regions with OM (e.g., Williams et al., 2014; Schaefer et al., 2014). However, Ruesch et al. (2014b) argue that this parameter is not uniquely diagnostic for impact-melt characterization.

To better understand the nature of compositional heterogeneity across Vesta, we undertake a comprehensive study of the spectral reflectance properties of HEDs and other types of materials known or inferred to be present on Vesta.

**2.0 Samples and laboratory spectra**

Reflectance spectra of two lines of mixtures between howardite and olivine or CM material are measured to understand how mixtures of Vesta-relevant materials vary spectrally. One mixing line is based on the diogenite-dominated howardite DaG 779, the other is based on the eucrite-dominated howardite NWA 1929. The carbonaceous chondrite material is from a specimen of the Jbilet-Winselwan CM2 carbonaceous chondrite. The olivine material is prepared from a forsterite-rich gemstone quality sample. Descriptions of the meteorites/samples are given in Appendix 1 with their compositional data in Appendix 2.

The meteorite samples are received in chips/bulks. The outer rims of each specimen are removed thoroughly before crushing to minimize the inclusion of fusion crust and terrestrial weathering products. They are ground to powders using an agate mortar either in a grinding mill or by hand at the Clausthal University of Technology, Germany (TUC). The fine-grained material are dry-sieved and further processed (milled) until all material passed through a 63 µm sieve. The sieved powders of the two howardites are used to prepare intimate mixtures in 10 wt.% intervals, each with olivine or Jbilet-Winselwan.

Individual powders are homogenized by stirring and shaking the powders in glass vials for about 1 minute before preparing the mixtures. The samples are mixed to weigh 250 mg in total using an analytical weighing balance in a clean laboratory at Max-Planck Institute for Solar System Research, Germany (MPS). Each mixture is again gently stirred for about 5 minutes using a pestle in a small agate mortar to produce homogeneous mixtures. The reflectance spectra (0.35-2.5 µm) of the particulate samples and their mixtures are measured at HOSERLab (University of Winnipeg, Canada) using an Analytical Spectral Devices (ASD) FieldSpec Pro HR spectrometer. The measurements are done relative to a Spectralon® standard using a 150W quartz-tungsten-halogen collimated light source. In each case, 200 spectra are acquired and averaged to provide sufficient signal-to-noise. The details of these measurements are discussed in Cloutis et al. (2013).

Apart from the measured spectra, available reflectance spectra of powdered HED meteorites, olivine, CM2, and olivine-orthopyroxene mixtures from the RELAB, USGS, and HOSERLab spectral libraries are used. Details on the spectra/samples are listed in Appendix 3. Figure 1 shows absolute (Fig. 1A) and normalized (Fig. 1B) reflectance spectra of the measured spectra (solid lines) of olivine, NWA 1929, DAG 779 and Jibilet Winselwan CM2. The Macibini-glass spectrum is from RELAB. The resampled data points at the effective wavelength of each filter are represented by the symbols in Fig. 1. The spectra are resampled to FC bandpasses using the instrument response function per filter (Sierks et al., 2011).

**3.0 Band Parameters**

The robustness of FC color data in constraining Vestan composition and mineralogy has been investigated in a number of studies. Despite the fact that the FC data does not cover the entire 1-µm absorption feature, mafic mineral compositions can still be derived and linked to particular HED lithologies (e.g., Reddy et al., 2012b, c; Thangjam et al., 2013, 2014; Nathues et al., 2014, 2015). Various FC band parameters have been developed to analyze and map lithologic units (Reddy et al., 2012; Thangjam et al., 2013), dark material (Reddy et al., 2012c; Nathues et al., 2014), olivine-rich regions (Thangjam et al., 2014; Nathues et al., 2015), and impact-melt/glass/orange material (Le Corre et al., 2013). Band parameters used in this study are listed in Table 1.

The spectral slope in the visible wavelength range of FC color data, generally defined by a ratio of reflectance values at 0.43 and 0.75 µm (0.43/0.75 µm), has been used elsewhere (e.g., Reddy et al., 2012b; Buratti et al., 2013; Le Corre et al., 2013; Schaefer et al., 2014) to analyze composition and surface properties of Vesta. The spectral slope has also widely implemented in comparative studies of Vesta, Vestoids, and HED meteorites (e.g., Buratti et al., 2013; Hicks et al., 2014; Hiroi and Pieters, 1998; Burbine et al., 2001). Computation of the slope from spectra normalized to a particular wavelength is typical for ground-based telescopic data (e.g., Luu and Jewitt, 1990; Bus and Binzel, 2002; Hicks et al., 2014). The general FC visible slope definition is similar to what has been applied to lunar Clementine Ultraviolet/Visible Camera (UVVIS) data (e.g., Pieters et al., 2001; Isaacson and Pieters, 2009). However, Vesta's regolith differs in many ways from the lunar regolith: for example, different space weathering processes (Pieters et al., 2012), absence/insignificance of lunar-like agglutinates and Fe°-np (nano-phase iron) and $TiO_2$ in HED meteorites (e.g., Buchanan et al., 2000; Singerling et al., 2013), dominantly mafic composition on Vesta unlike the prominent lunar anorthositic (felsic) highlands and mafic mare, etc. Again, the FC is equipped with 4 filters (0.43, 0.55, 0.65, 0.75 µm) in the visible wavelength range compared to 2 filters (0.41, 0.75 µm) for the Clementine UVVIS camera. We define visible slope (VS) with slight modifications from Luu and Jewitt (1990), Doressoundiram et al. (2008), and Hiroi and Pieters (1998). The slope is calculated in % per 0.1 µm from the absolute reflectance spectra, over a wavelength range framed by minimum reflectance and a maximum reflectance in the visible wavelength range.

Band strength or band depth of the 1-µm absorption feature is generally represented by a ratio of the reflectance values at 0.75 and 0.92 µm (0.75/0.92 µm) for FC color data (Reddy et al., 2012a; Le Corre et al., 2013; Thangjam et al., 2014). This is similar to the lunar Clementine parameter (e.g., Tompkins and Pieters 1999; Pieters et al. 2001; Isaacson and Pieters 2009). It is worth mentioning that Tompkins and Pieters (1999) defined this parameter by a ratio of 0.75 µm to any of the bands near 1 µm (0.75/0.90 µm, 0.75/0.95 µm, 0.75/1.0 µm). They used one of the ratios, depending on the closest wavelength to the 1-µm absorption band minimum, to estimate a first-order relative abundance of mafic minerals (and plagioclase). Though this parameter is diagnostic to analyze lunar surface's FeO abundance and soil maturity and their composition, it is important to be cautious in applying this parameter to Vesta's surface because of the aforementioned differences between the two bodies. Here, we define the parameter Band Strength (BS) as the ratio of maximum reflectance value in the visible wavelength range ($\lambda_{max\_VIS}$) to the minimum reflectance in the near-infrared range ($\lambda_{min\_NIR}$) (see Fig. 1C).

Band Tilt (BT) is one of the most diagnostic parameters in the near-infrared region to characterize the 1-µm absorption feature of mafic minerals. This is discussed in detail in Thangjam et al. (2013, 2014), and Nathues et al. (2015). This parameter is basically a lunar Clementine UVVIS parameter, which was introduced by Pieters et al. (2001), and then later modified by Dhingra (2008), and Isaacson and Pieters (2009).

Mid Ratio (MR) was introduced in Thangjam et al. (2014) and discussed in detail elsewhere (Thangjam et al., 2014; Nathues et al., 2015). This parameter is useful for distinguishing olivine-rich lithologies from HED lithologies, especially from High-Ca pyroxenes observed in HEDs (Thangjam et al., 2014).

A sketch illustrates our band parameters together with a typical HED spectrum (Fig. 1C). The wavelengths with minimum reflectance ($\lambda_{min\_VIS}$) and maximum reflectance ($\lambda_{max\_VIS}$) in the visible wavelength range, and minimum reflectance in the near-infrared region ($\lambda_{min\_NIR}$) are determined before calculating the band parameters. The band parameter values of the spectra used in this study are provided in Appendix 3.

## 4.0 Analysis - Laboratory spectra

More than 80% of eucrites and diogenites show $\lambda_{max\_VIS}$ at 0.75 and 0.65 μm, respectively. The spectrum of eucrite-dominated howardite NWA 1929 shows $\lambda_{max\_VIS}$ at 0.75 μm, while $\lambda_{max\_VIS}$ of the diogenite-dominated howardite DaG 779 spectrum is at 0.65 μm. Both the olivine and the Macibini-eucrite glass spectra exhibit $\lambda_{max\_VIS}$ at 0.55 μm. The spectrum with the steepest visible slope is the Macibini-eucrite glass, followed by NWA 1929, olivine, and DaG 779. Similarly, the glass spectrum shows the deepest band strength, followed by DaG 779, NWA 1929, and olivine. The Jbilet-Winselwan spectrum appears to be almost flat in the FC wavelength range though it displays a notably visible slope.

Figure 2 presents FC resampled spectra of Jbilet-Winselwan, howardites DaG 779, and NWA 1929, and their mixtures. Absolute reflectance spectra of these mixtures are shown in 10 wt.% intervals (Fig. 2A and 2B) whereas normalized spectra are displayed in steps of 10, 30, 50, 70 and 90 wt.% for clarity (Fig. 2C and 2D). The two howardites show significantly higher reflectance and band strength values compared to that of Jbilet-Winselwan. Both the parameter values are significantly reduced by adding 10 wt.% of Jbilet-Winselwan, while alternatively, rather a large amount of howardites (more than 30-40 wt.%) can increase the low reflectance and band strength values of Jbilet-Winselwan. Though a rapid change of reflectance and band strength is apparent in these mixtures (Fig. 2A, B), only slight variations are observed in the visible slope (Fig. 2C, D). The visible slope values of mixtures with DaG 779 increase with increasing CM2 abundance, whereas, in case of NWA 1929, this behavior is less systematic. This is probably due to differences in the reflectance in the visible wavelength range of the different samples that affects their spectral slope. The majority of the mixtures show $\lambda_{max\_VIS}$ at 0.75 μm, and the spectral slope from 0.43 μm to either 0.65 or 0.55 µm drops very rapidly compared to the spectral slope from 0.65 to 0.75 µm.

Figure 3 shows resampled spectra of mixtures of olivine and howardite (either with DaG 779 or NWA 1929). Absolute reflectance spectra in 10 wt.% intervals (Fig. 3A, B) and normalized spectra at steps of 10, 30, 50, 70, 90 wt.% howardite content (Fig. 3C, D) are displayed. Changes in the spectral shape and their band parameters, such as visible slope, band tilt, and band strength are noticed with changes in end member abundances. Unlike the CM2 spectrum, the olivine spectrum exhibits a notable 1-µm feature. The olivine spectrum shows higher reflectance compared to the howardites. Visible slope and reflectance of mixtures gradually changes as a function of the components. The same is true for parameters BT and BS. Olivine shows higher values of BT but lower BS parameters than the two howardites. A significant change in reflectance, VS, BT and MR values of olivine is seen by adding even only 10 wt.% howardite. A howarditic amount of more than 30-40 wt.% in these mixtures produces spectra similar to HEDs. This is almost opposite to what we observed in CM2 mixtures where the howardites spectra are significantly affected by even 10 wt.% of CM2.

Eucrite and olivine show larger BT values than diogenites, and therefore, the BT parameter is the most effective for distinguishing diogenite from eucrite and olivine (Thangjam et al., 2013, 2014). The MR values of diogenite, olivine, and olivine-rich olivine-orthopyroxene mixtures are in general larger than eucrite and clinopyroxene found in HEDs (Thangjam et al., 2014). The CM2s spectra show much lower reflectances compared to HEDs and olivine, and thus, the reflectance serves a useful criterion to recognize the presence of CM2 material (Fig. 4). Macibini-eucrite glass spectrum displays a rather large value of VS compared to the HEDs, olivine, and CM2s, and hence, the VS parameter can be useful to identify and discriminate the glass spectrum (Fig. 4). However, the VS values of HEDs, olivine, and CM2s overlap. Similarly, the BS parameter shows a broad range of values for HEDs and olivine, though the BS parameter is generally used to assess abundance of mafic minerals (e.g., Reddy et al., 2012b; Thangjam et al., 2013). This indicates that multiple spectral parameters are required for more robust discrimination of the various materials included in this study.

**4.1 Two-dimensional approach**

Two-dimensional approaches (i.e., two band parameter spaces) are generally employed in earlier works to characterize surface lithologies, for example, Band Tilt versus Band Curvature parameter space to distinguish among eucrites and diogenites (Thangjam et al., 2013), and Band Tilt versus Mid Ratio to separate olivine from HEDs (Thangjam et al., 2014; Nathues et al., 2015), etc. Apart from the aforementioned band parameter spaces, other parameter spaces such as band strength or visible slope versus reflectance at 0.75μm need to be discussed, since they are suggested to be diagnostic for identifying impact melts, exogenic dark materials, etc., (e.g., Le Corre et al., 2013; Reddy et al., 2012b; McCord et al., 2012). A detailed analysis of these parameter spaces is presented here.

The parameters VS and BS are plotted versus reflectance at 0.55 μm ($R_{0.55\mu m}$) for Macibini-eucrite glass, howardites DaG 779 and NWA 1929, olivine, Jbilet-Winselwan, and mixtures of howardites with olivine or CM2 (Fig 4). In the VS versus $R_{0.55\mu m}$ band parameter space (Fig. 4A), the Jbilet-Winselwan, olivine, and glass sample are well separated from each other and the two howardites. The CM2-chondrite mixtures and the olivine mixtures plot on either side of the howardite samples along the x-axis ($R_{0.55\mu m}$). We do not have data points of mixtures of glass and howardite, but a trend more or less in between the howardites and the glass can be expected. The parameter VS of the olivine and howardite mixtures show similar values with a negligible change compared with the howardites though the values are significantly different from the nearly pure olivine. The samples and mixtures in Fig. 4A are not distinguishable if all the eucrite, diogenite, olivine, and CM2 are plotted together (Fig. 4C). Similar plots are displayed for BS versus $R_{0.55\mu m}$ (Fig. 4B, D). The BS parameter of Macibini-glass is not as distinct as for the VS parameter, and this data point is not separable from the eucrite, diogenite, and olivine. These analyzes suggest that the VS and BS parameters are not ideally suited to distinguish lithologies among HEDs, olivine, and their mixtures. Meanwhile, visible slope and band strength parameters are known to be significantly affected not only by the mineralogy but multiple parameters such as grain size, viewing geometry or phase angle, temperature, etc. (e.g., Nathues et al., 2000; Reddy et al., 2012c; Duffard et al., 2005; Izawa et al., in review; Cloutis et al., 2013; Moskovitz et al., 2010).

Though the two-dimensional parameter spaces are useful to identify particular lithologies of interest, they seem to be insufficient in dealing with multiple components. Therefore, an advanced spectral approach of three-dimensional analysis (i.e., three-band parameter space with polyhedrons) is introduced and applied for the first time in this study. It is worth mentioning that the three-dimensional parameter space used elsewhere (e.g., Filacchione et al., 2012) are scattered data points, however the polyhedrons defined in this study allow to assess both qualitative and quantitative information. Again, this approach is different from other methodologies employing multi-dimensional parameters defined through statistical analyses for example, principal component analysis (e.g., Roig and Gill-Hutton, 2006; Nathues et al., 2012), or cluster analysis (e.g., Pinilia-Alonso et al., 2011), etc.

## 4.2 Three-dimensional approach

A three-dimensional band parameter space uses a further parameter (information), and thus, the results can be more robust and precise. The band parameter values computed from laboratory spectra representing different lithologies are used to define a three-dimensional space spitted in several convex polyhedrons. The polyhedrons are constructed by many convex triangular facets, using the band parameter values as vertices and a connectivity array defined by the convex hull method. Plotting are done in IDL programming language. Various sets of combinations of the band parameters are discussed, and a particular parameter space of interest is chosen for further analysis and application.

The polyhedrons of the BT versus MR versus $R_{0.55\mu m}$ band parameter space are shown in Fig. 5A in two different views. The parameters BT, MR, and $R_{0.55\mu m}$ are plotted along the X, Y, and Z axis, respectively. Each polyhedron or data point is given a name for clarity (Euc: eucrite; Dio: diogenite; How: howardite; Ol: olivine; Ol-HED: olivine plus howardite mixtures, and olivine-bearing diogenites; Ol-rich-Opx: Olivine rich olivine-orthopyroxene mixtures (>40 wt% olivine); CM2: CM2 chondrite; CM2-HED: Jbilet-Winselwan plus howardite mixture, Murchison plus eucrite mixture, and CM2-rich howardite; Gl: Macibini eucrite glass). The polyhedrons of diogenite and eucrite are clearly distinct while the howardite polyhedron overlaps with both of them. The polyhedrons of olivine and CM2-chondrites are separable from HEDs while the CM2-HEDs and Ol-HEDs polyhedrons gradually merge with the howardite/eucrite polyhedron. The CM2-HEDs polyhedron overlaps with that of eucrite/howardite at a range of approximately 20-30 wt.% CM2 content, whereas the overlapping is at about 40-60 wt.% of olivine content for the Ol-HEDs polyhedron. The data point of the glass neither overlaps with any polyhedron of olivine nor Ol-HEDs, but it lies within the Ol-rich-Opx and Ol-HEDs polyhedrons. Thus, the glass sample is not uniquely distinguishable in this band parameter space.

Two different views of the polyhedrons in the BT versus MR versus VS band parameter space are shown in Fig. 5B. Diogenite, howardite, olivine, CM2, CM2-HEDs, Ol-HEDs display a broad range of VS values, and they overlap to varying extents with each other, making it difficult to distinguish them. However, CM2, olivine, Ol-rich-Opx mixtures are distinguishable from the HEDs. Eucrite and diogenite are separated from each other but overlap with howardite. Since the data point of Macibini-glass shows a rather large VS value, this data point is distinct. Thus, this parameter space can be used to identify eucrite-glass samples assuming that glasses in HEDs are similar to the Macibini eucrite glass. However, data points of two Padvarninkai eucrite impact melts, and a heavily shocked eucrite, JaH 626, do not follow the glass sample, and they overlap with other polyhedrons. A small amount (< 10-20 wt.%) of howardite mixed with glass can yield spectra similar to HEDs (e.g., Buchanan et al, 2014), and it may further complicate separation of such data points from HEDs.

The polyhedrons of the BT versus MR versus BS band parameter space are shown from two different views in Fig. 5C. Some of the polyhedrons are distinct, and the BS parameter is insensitive to distinguish among HEDs, olivine, and Ol-HEDs. Eucrite and diogenite are separable from each other. Olivine, CM2, and Ol-rich-Opx polyhedrons are distinguished from HEDs. The Ol-HEDs and CM2-HEDs polyhedrons overlap with the HEDs to a large extent. The data point of the eucrite glass sample does not overlap with either of the polyhedrons though it lies close to the polyhedrons of olivine and Ol-rich-Opx. However, impact-melt and shock HED samples plot far apart from the glass sample and are indistinguishable from other polyhedrons.

It is rather difficult to distinguish the various lithologies in the VS versus BS versus $R_{0.55\mu m}$ band parameter space (Fig. 5D). The CM2 and the Macibini eucrite glass sample are offset from the HEDs. The glass sample does not overlap with the HEDs. However the other impact-melt and shock HED samples neither follow the glass sample nor separable from HEDs.

**5.0 Implications for Dawn FC at Vesta**

The three-dimensional polyhedrons defined from various laboratory spectra are applied to Dawn FC data to enable compositional analysis and mapping. Algorithm in IDL program has been employed to assign FC data points to the respective polyhedrons. Details of the algorithm are provided in Appendix 4.

As a case study, we select Arruntia region in the northern hemisphere of Vesta. This region shows many interesting lithologies, including olivine-rich sites (Ammannito et al., 2013a; Thangjam et al., 2014; Ruesch et al., 2014b; Nathues et al., 2015), impact-melt/glass/shocked or orange material (Le Corre et al., 2013), and dark and bright material (e.g., Thangjam et al., 2014; Ruesch et al., 2014a; Zambon et al., 2014). The Arruntia crater is described as one of the freshest impact craters on Vesta, showing dark and bright ejecta (Ruesch et al., 2014b). We use FC color data (~ 60 m/pixel) obtained during High Altitude Mapping Orbit (HAMO and HAMO 2) for spectral analysis and mapping. The calibration and processing of the FC color data including photometric corrections and error or uncertainty in the data are discussed in Nathues et al. (2014). The errors in the spectral data points are usually less than the size of the symbols displayed in the plots. For investigations of morphological features, the FC clear filter images (~ 18 m/pixel) obtained during Low Altitude Mapping Orbit (LAMO) is used. Figure 6 shows the reflectance and several band parameter images of our study area: (A) reflectance at 0.55 µm ($R_{0.55\mu m}$), (B) MR (Mid Ratio), (C) VS (Visible Slope in % per 0.1 µm), (D) BS (Band Strength), and (E) BT (Band Tilt). Gray material (mean $R_{0.55\mu m}$ ~ 0.19) covers the region predominantly, while bright material (mean $R_{0.55\mu m}$ ~ 0.25) is more abundant than the dark material (mean $R_{0.55\mu m}$ ~ 0.14).

Bright material is found in the ejecta and on the wall of the crater, whereas dark material is locally enriched in a few regions of ejecta and the crater wall. The band parameter values (VS, BS, BT, and MR) do not follow a clear trend of dark, bright and gray material, but the majority of the bright material in the ejecta usually exhibits higher BT, MR, VS values and lower BS values. A slope map is derived from the HAMO-DTM (~ 62 m/pixel) to visualize the relief of this region (Fig. 6F). The Arruntia crater walls show rather steep slopes compared to other craters in this region. The topographic profile of this crater resembles a V-shape without having a pronounced flat crater floor. The distribution of the ejecta material follows the local relief.

For our application, we choose the three-dimensional polyhedrons defined by the BT, MR and $R_{0.55\mu m}$ parameters. The band parameter values of the inflight data are calculated using an IDL algorithm similar to that used for laboratory spectra. Areas with the dominant lithology in this region, i.e., eucrite (Fig. 7A) are mapped. Figure 7B shows those areas (~ 23%) that are assigned to the CM2-HEDs polyhedron where less than 20-30 wt.% of CM2 content is indistinguishable from eucrite/howardite. However, the CM2-rich localities ($\geq$ 20-30 wt.% of CM2, distinguished from eucrite/howardite) are very minor in extent (< 1% of the total area, figure not shown), and this suggests that dark material, particularly CM2, does not contribute to this region significantly. The mapping of lithologies containing olivine mixtures with HEDs is accomplished by a combined polyhedron of Ol-HEDs and Ol-rich-Opx. The Ol-rich-Opx mixtures (>40 wt.% olivine, Appendix 3) are already used to identify and map olivine-rich lithologies by Thangjam et al. (2014) and Nathues et al. (2015). Figure 7C shows areas of olivine mixed with HEDs where less than 40-60 wt.% of olivine content are not separable from eucrite/howardite. There is no unambiguous interpretation for this observation because they can be either eucrite/howardite without olivine or below detection limits. Figure 7D shows olivine-rich areas ($\geq$ 40-60%) that are distinguished from eucrite/howardite. These potential and distinct olivine-rich areas are consistent with the olivine-rich exposures identified by Thangjam et al. (2014) but are more abundant. These exposures cover approximately 4% of the study region. Within a region of 2.5 crater radii, the olivine-rich sites occupy about 14% of the surface, which is 12% more than what Thangjam et al. (2014) found. Most of the olivine-rich sites are on the ejecta near the crater rim, with few sites on the crater wall and the floor. Based on our observations, eucrite covers the majority of the Arruntia region, and therefore, this implies that this region is dominantly eucrite-rich howardite in composition, which is consistent with earlier observations (Thangjam et al., 2014; Ammannito et al., 2013a). Most probably there are no exposures of (nearly) pure diogenite, olivine or CM2-material in this region. The absence of diogenite was also noted by Ammannito et al. (2013a) and Thangjam et al. (2014). We are not able to identify sites consisting mainly of glass or impact-melts because of the paucity of laboratory impact-melt/glass spectral data. Meanwhile, we do not find any visible slope value that is comparable to the Macibini-glass sample.

**6.0 Discussion**

**6.1 Olivine on Vesta**

Olivine on Vesta is currently an unresolved issue, but crucial for understanding the evolutionary aspects (e.g., Clenet et al., 2014; Consolmagno et al., 2015). Despite the predictions of olivine in the Rheasilvia basin (e.g., Gaffey, 1997; Thomas et al., 1997) and an olivine-rich mantle of Vesta (e.g., Righter and Drake, 1997), Dawn observations have not unambiguously detect olivine of mantle origin yet. Most of the olivine-rich sites are detected in the northern hemisphere instead (e.g., Thangjam et al., 2014; Nathues et al., 2015). Because of the lack of significant amount of olivine in the Rheasilvia basin, Clenet et al. (2014) suggest a deeper crust-mantle transition zone of Vesta (up to 80-100 km) which is quite unusual compared to other terrestrial silicate-dominated bodies. The unresolved issue of olivine is one of the arguments by Consolmagno et al. (2015) who claimed that Vesta is probably not an intact and pristine body.

Based on linear spectral unmixing analysis, Combe et al. (2015) argue that the olivine-rich areas identified at Bellicia, Arruntia and Pomponia craters are possible without olivine, i.e., a mixture of low- and high-Ca pyroxenes (hypersthene, pigeonite and diopside). However, the presence of olivine-rich material in the Arruntia region is obvious based on our three-dimensional spectral analysis as well as the earlier works (Ammannito et al., 2013a; Thangjam et al., 2014; Nathues et al., 2015; Ruesch et al., 2014b; Palomba et al., 2015). Spectral modeling by Poulet et al. (2014) suggests that olivine mixed with howardite is likely a ubiquitous component on Vesta. Nathues et al. (2015) and Le Corre et al. (2015) mention that olivine on Vesta is exogenic in origin. However, Cheek and Sunshine (2014) claim that olivine in Bellicia and Arruntia region are of shallow crustal plutonic in origin because they observe localities where Cr-rich pyroxene and olivine occur together. They further mention that the presence of such differentiated plutons on Vesta favor the evolution model of late stage serial magmatism (e.g., Yamaguchi et al., 1997; Mittlefehldt, 1994). However, there is geochemical and microstructural evidence who claimed the presence of potential olivine-bearing mantle material in few olivine-rich diogenites (Tkalcec et al., 2013; Tkalcec and Brenker, 2014) and howardites (Lunning et al., 2015; Hahn et al., 2015). Hahn et al. (2015) claim that they discovered harzburgite clasts of Mg-rich pyroxene plus olivine in howardites that preserved recrystallized textures. These authors hypothesize that the upper mantle of Vesta is harzburgitic in composition.

Thus, the origin of olivine on Vesta is still under debate. Meanwhile, limitations and difficulties in detecting olivine on Vesta's surface is also of concern, because the spectral signature of olivine may be either masked by the presence of pyroxenes (e.g., Beck et al., 2013) or is spatially unresolved by the instruments onboard Dawn (e.g., Beck et al., 2013; Jutzi et al., 2013). Our spectral analysis technique still leave open the possibility of up to a few tens of weight percent olivine to be present on Vesta that is spectrally undetectable.

**6.2 'Orange material' or shocked/glass/impact-melt**

Le Corre et al. (2013) identified sites of 'Orange Material' (OM) mostly in equatorial and northern hemispheric regions including the Arruntia region. They suggest that such material shows red slope in the visible wavelength range. However, visible slope that is comparable with the Macibini-eucrite glass is not found in this work, though it is not necessarily true that all HED glass spectra should show similar characteristics to this sample. Eucrites with significant glass/impact melt components (e.g., LEW 85303, Padvarninkai) are neither distinguishable from the general eucrites nor match the visible slope with the nearly pure Macibini-eucrite glass. On the other hand, other components like olivine do exhibit red visible slope. Therefore, a red spectral slope in the visible wavelength range is not necessarily a unique diagnostic parameter of impact-melt-bearing HEDs. This point is noted by Ruesch et al. (2014b), too. From the morphological observations in the Arruntia crater, Ruesch et al. (2014a) suggest the possibility of the presence of thin impact melt veneers in the ejecta. We investigated morphological features for impact melts in and outside Arruntia using high-resolution clear filter LAMO images (~ 18 m/pixel). Potential impact-melt flow features are found on the crater wall, floor and ejecta (Fig. 8). The HAMO image (~ 60 m/pixel) shows the locations (Fig. 8A) of the putative impact-melt features identified in LAMO images (Fig. 8B-F). Streaks of dark material are observed in the continuous ejecta near the crater rim (Fig. 8B) and also on the inner crater wall. Streaks of dark material on the rim and ejecta (away from the crater) coated with thin veneers of impact melt are one of the common lunar impact-melt features, particularly in small lunar highland craters (e.g., Plescia and Cintala, 2012). The available resolution is not sufficient to resolve the features in fine details as was inspected from Lunar Reconnaissance Orbiter Camera - Narrow Angle Camera (e.g., Plescia and Cintala, 2012). Though the features are difficult to distinguish from ejecta debris, the dark streaks may represent pockets of impact melts ejected during the impact. Such streaks are found mainly in the continuous ejecta near the rim along with the bright material. More typical impact melt flow features are detected on crater wall slopes (Fig. 8C, D). The feature on the northwest flank of the crater wall (Fig. 8C) shows what appear to be pressure ridges and prominent levees. Pressure ridges are seen in lunar impact-melt flows where such features occurred in late stage melt-flows (e.g., Carter et al., 2012). Such ridges are also frequently observed in terrestrial lava flows, where viscous material is compressed due to horizontal and vertical velocity gradients and often develop at distal ends (Fink and Fletcher, 1978; Theilig and Greeley, 1986). Another impact-melt flow feature is observed at the northeastern wall showing faint pressure ridges and prominent levees (Fig. 8D). Both the features (Fig. 8C, D) occur along the steep slopes of the walls, and their widths become narrower and merge downslope as the underlying slope becomes shallower towards the floor. The floor of Arruntia crater is partially in the shade, but weakly illuminated by multi-scattered light, and therefore, it is difficult to inspect the features. The visibility of the floor of the crater has been enhanced in this scene (Fig. 8E). Bright-toned debris covers the floor, but the smooth-textured and comparatively darker material seems to overlie the debris partly. The flow features are not only detected on the floor and the ejecta near the rim, but are also found on the outer rim and in the discontinuous ejecta of the crater (Fig. 8F). The lobate flow features that exhibit a smooth texture and significant difference in spatial distribution of craters represent a potential impact melt veneer deposited on rather a densely-cratered surface. These features extend as continuous and discontinuous lobes. Similar putative impact-melt flow features are identified on Vesta by Williams et al. (2014). Based on numerical modeling and scaling laws, Williams et al. (2014) suggest that impact-melt formation on Vesta is possible though the volumes are less compared with similar-sized lunar craters. On the other hand, Keil et al. (1997) claim that impact-melt deposits on asteroids like Vesta are negligible because of rather low impact velocities compared to the terrestrial planets. It is worth mentioning that impact-melt/shock/glass has been identified in HED meteorites but is not as common and abundant as in lunar and Martian achondrites (e.g., Buchanan et al., 2000; Singerling et al., 2013; Rubin, 2015). Material like impact-melts, shocked- and glassy-components in HEDs may bear different compositional/mineralogical characteristics depending on the intensity of the shocked pressures/temperatures or thermal metamorphism (e.g., Rubin, 2015; Takeda and Graham, 1991), and accordingly they are expected to show different spectral features.

## 6.3 Dark and bright material

Vesta is known for showing significant albedo variations across its surface (e.g., Reddy et al., 2012b, c). Dark material is more abundant than bright material (e.g., Palomba et al., 2014; Zambon et al., 2014). Palomba et al. (2014) suggest that dark material is dominantly eucritic or howarditic in composition with significant contributions of exogenous carbonaceous chondritic material. They mentioned that shocked components in HEDs could also be a representative of dark material on Vesta because of their low reflectances. Bright material is interpreted as relatively fresh and recently excavated sub-surface components because of their higher reflectance and deeper band depth values (Zambon et al., 2014). Olivine-rich areas in Bellicia and Arruntia region are noted as a constituent of the bright material (Zambon et al., 2014). They note that the bright material units of Arruntia region are mixed with dark material. However, Palomba et al. (2014) does not mention the presence of dark material in and around the Arruntia region though they defined at least 15% lower reflectance than the local average as one of the criteria to identify this material. For Arruntia region, we observe that dark, bright and gray regions show average reflectance ($R_{0.55\mu m}$) of approximately 0.14, 0.19, and 0.25, respectively. The variation of reflectance of the gray material with the bright and dark material are about 31% and 26%, respectively, while the variation of bright to the dark material is about 66%. Many of the bright areas on the inner walls, and few ejecta of the crater display rather a high reflectance (~ 0.4 or more). The CM2-rich exposures (> 20-30% of CM2 content) are found in very few localized areas. The existence of plenty of areas with lesser amounts of CM2-material (less than 20-30 wt.%) is possible, mixed with eucrites/howardites (Fig. 7B). The olivine-rich and possibly olivine-poor areas also display higher reflectance than the surrounding areas (Fig. 7C, D).

**7.0 Summary and future work**

We introduce and apply an innovative three-dimensional spectral approach (three band parameter space with polyhedrons) that better characterizes the heterogeneous surface compositions compared to the generally used two-dimensional spectral approach (two band parameter space). Several lithologic units are mapped using the FC color data based on the laboratory polyhedrons defined from various spectra of HEDs, olivine, CM2, and their mixtures. According to this study, we find that eucrite-rich howardite is the dominant composition in the Arruntia region. Olivine-rich exposures are found in rather large spatial extents compared to the earlier studies. Spectral features in FC color data do not allow unique identification of impact-melt/shock/glass materials; however, morphological features investigated using higher resolution FC clear filter images identified potential impact melt flow features in this region. Nearly pure CM2, olivine, glass and diogenite exposures are likely not present in this region. This three-dimensional spectral analysis can be applied to the entire Vestan surface to map the compositional heterogeneity. The applicability of this approach to the hyperspectral datasets such as the VIR data as well as to other planetary bodies are yet to be proven.

**Acknowledgements**

We are thankful to the RELAB and USGS spectral library for the spectra used in this work. G.T. is grateful to Ladislav Rezac, Nagaraju Krishnappa, Jayant Joshi, Dick Jackson and Megha Bhatt for their help in IDL/MATLAB/python programming. G.T. would like to heartily thank many colleagues from TU/Clausthal (Cornelia Ambrosi, Silke Schlenczek, Dietlind Nordhausen), TU/Hannover (Harald Behrens), and MPS (Walter Goetz, Harald Steininger, Henning Fischer) for their help during the preparation of the samples/mixtures. G.T. is thankful to Pierre Beck for providing spectra of olivine-rich diogenites NWA 5480 and NWA 4223. E.A.C. thanks the Canadian Space Agency, the Canada Foundation for Innovation, the Manitoba Research Innovations Fund, and NSERC for supporting the establishment and operation of HOSERLab at the University of Winnipeg. We would like to acknowledge the two anonymous reviewers for their thorough review and constructive comments that helped to improve the manuscript.

**Appendix 1: Description of the meteorite specimens used in this study**

The meteorite Dar al Gani 779 (DaG 779) has been classified as a howardite with a weathering grade W1 and a shock stage S2 (Grossman, 2000). The modal mineralogy is calculated by using the mineral data (electron microprobe) and whole rock XRF analyzes at TUC. Optical inspection in plane and crossed polarized light reveals distinct clasts in a fine-grained matrix. The majority of the clasts are of diogenite composition, which is dominated by orthopyroxene or, low calcium pyroxene. In size, they are not larger than 5 mm. Few lithic clasts of eucritic composition are present. Mineral analyzes show non-uniform composition of pyroxenes. Their composition ranges from a more enstatitic (~$Wo_{4\pm3}En_{66\pm5}Fs_{30\pm5}$) to pigeonite (~$Wo_{7\pm1}En_{36\pm2}Fs_{57\pm2}$). With increasing Ca-content, there is augite (~$Wo_{43\pm1}En_{29\pm1}Fs_{29\pm2}$) as well as rare diopsides. Plagioclase feldspars range from $An_{70}$ to $An_{94}$ with negligible K-feldspar component. Minor minerals are spinel (mostly chromite in composition), ilmenite, olivine (~$Fo_{67}$), a silica phase, troilite and rare Ca-phosphate. This howardite contains 77 wt.% orthopyroxene, 8 wt.% plagioclase, 7 wt.% clinopyroxene, 2 wt.% spinel and minor content (≤1 wt.%) of olivine, silica phase, troilite, ilmenite, apatite/whitlockite and rare terrestrial weathering products, such as Ca-carbonates and iron-oxides/-oxy-hydroxides. In DaG 779, a number of secondary processes typical for howardites are observed: 1) recrystallization after melting due to an impact, especially in eucritic clasts; 2) rapidly cooled glasses with quench crystals; 3) metasomatic alteration reactions producing secondary augite, a silica phase, and FeS from pigeonite plus sulfur-rich fluid.

The meteorite North West Africa 1929 (NWA 1929) has been classified as a howardite (Connolly et al., 2006). They find that this specimen consists of 72 vol.-% cumulate eucrite clasts ($Fs_{45-40}Wo_{7-20}$; plagioclase: $An_{91.2-95.3}$; metal: Ni = 0.97 wt.-%, Cr = 0.87 wt.-%), 14 vol.-% diogenite clasts (pyroxene: $Fs_{43-54}Wo_{2.5-3.6}$; Fe/Mn = 36, 37), 6 vol.-% melt (glass) clasts, and 8 vol.-% subophitic clasts. The eucritic clasts look like coarse-grained gabbroic rocks (Connolly et al., 2006). Solid state recrystallization of pyroxene and plagioclase occurs throughout the sample, showing localized melt pockets/veins within clasts (Connolly et al., 2006). NWA 1929 yields a bulk density of 2.91 g/cm$^3$ (McCausland and Flemming, 2006). Howardite NWA 2698 and eucrite NWA 2690 are probably paired with NWA 1929 (Connolly et al., 2006).

The chondrite Jbilet-Winselwan is categorized as a carbonaceous meteorite, Type II (CM2) with shock stage S0 and weathering grade W1 (Meteoritical Bulletin 102, in preparation). Hewins and Garvie (Meteoritical Bulletin 102, in preparation) find that the specimen they analyzed contains chondrules/fragments of Types I and II that includes barred olivine-porphyritic olivine of formerly metal-rich and olivine-pyroxene Type I chondrules, and forsterite relict-grained Type II chondrules. They reported that most of the chondrules sizes are around 200 μm, but range up to 1.2 mm, while a few CAIs are 800 μm in size. They also noted that powder X-ray diffraction analysis show serpentine, smectites, and tochilinite. Hewins (Meteoritical Bulletin 102, in preparation) analyzed olivine composition ranging from $Fa_{0.98\pm0.44}$ ($Fa_{25-40}$) to $Fa_{2.6\pm1.5}$ ($Fa_{40-61}$). Russell et al. (2014) report that Jbilet-Winselwan experienced only minor aqueous alteration though a few severely altered portions is also seen. They observe rather clear chondrules in a fine-grained matrix. Whole rock analyzes chemical data are not yet available.

Olivine used here for mixtures with howardite and CC is from the Mineral Collection of Clausthal University of Technology, their origin is unknown but probably from reworked mantle xenoliths similar to those of Navajo Co. AZ, USA deposits. Individual grains (3 to 7 mm in size) are of gem stone quality with very rare inclusions of chromite.

**Appendix 2:** Chemical whole rock data of two howardites and an olivine used for mixtures. The result of Jbilet-Winselwan is not yet available. Data from XRF based on fused glass diluted by $Li_2B_4O_7$ (1:5).

| Whole-rock composition | | | |
|---|---|---|---|
| **Wt. %** | **DaG 779** | **NWA 1929** | **Olivine** |
| $SiO_2$ | 47.79 | 46.22 | 38.92 |
| $TiO_2$ | 0.16 | 0.67 | 0.01 |
| $Al_2O_3$ | 3.46 | 11.56 | 0.02 |
| $Al_2O_3$ | 18.30 | 19.32 | 10.07 |
| **MnO** | 0.51 | 0.52 | 0.11 |
| **MgO** | 20.17 | 7.31 | 50.93 |
| **CaO** | 3.28 | 10.02 | 0.07 |
| $Na_2O$ | 0.13 | 0.43 | - |
| $K_2O$ | 0.02 | 0.05 | - |
| **Cr (ppm)** | 7438.00 | 13.00 | 100.00 |
| **Co (ppm)** | 19.00 | 4.00 | - |
| **Ni (ppm)** | 33.00 | 29.00 | 2700.00 |
| **∑ (%)** | 94.57 | 96.15 | 101.04 |

**Appendix 3:** Spectra/samples and band parameters (values) used in this study

| Euc | Size (µm) | BT | MR | R0.555 | BS | VS | Source |
|---|---|---|---|---|---|---|---|
| A-87272 | 0-25 | 0.97 | 0.89 | 0.49 | 2.09 | 5.74 | Relab |
| A-881819 | 0-25 | 0.93 | 0.88 | 0.48 | 1.82 | 5.14 | Relab |
| ALH-78132 | 0-25 | 0.92 | 0.86 | 0.37 | 1.63 | 4.16 | Relab |
| ALH-78132 | 25-45 | 0.91 | 0.87 | 0.22 | 2.15 | 3.40 | Relab |
| ALH-78132 | 45-75 | 0.91 | 0.91 | 0.17 | 2.33 | 2.74 | Relab |
| ALH85001 (dry-sieved) | 0-25 | 0.91 | 0.90 | 0.51 | 1.68 | 3.16 | Relab |
| ALH85001 (wet-sieved) | 0-25 | 0.92 | 0.89 | 0.42 | 1.56 | 2.32 | Relab |
| ALHA76005 | 0-25 | 0.95 | 0.85 | 0.43 | 1.52 | 4.23 | Relab |
| ALHA76005 | 125-250 | 0.96 | 0.92 | 0.18 | 1.78 | 1.74 | Relab |
| ALHA76005 | 250-500 | 0.97 | 0.90 | 0.18 | 1.65 | 1.45 | Relab |
| ALHA76005 | 25-45 | 0.94 | 0.85 | 0.29 | 1.79 | 3.75 | Relab |
| ALHA76005 | 45-75 | 0.94 | 0.85 | 0.21 | 1.99 | 2.91 | Relab |
| ALHA76005 | 75-125 | 0.95 | 0.89 | 0.19 | 1.88 | 2.21 | Relab |
| ALHA76005 | 0-250 | 0.96 | 0.89 | 0.23 | 1.66 | 2.37 | Relab |
| ALHA81001 | 0-125 | 0.93 | 0.81 | 0.20 | 1.49 | 1.20 | Relab |
| ALHA81001 | 0-45 | 0.95 | 0.84 | 0.33 | 1.35 | 2.58 | Relab |
| ALHA81011 | 0-125 | 1.00 | 0.86 | 0.24 | 1.71 | 1.69 | Relab |
| Bereba | 0-25 | 0.96 | 0.84 | 0.39 | 1.58 | 2.50 | Relab |
| Bouvante | 0-500 | 0.99 | 0.85 | 0.19 | 1.53 | 0.77 | Relab |
| Bouvante | 0-25 | 0.96 | 0.78 | 0.27 | 1.56 | 3.49 | Relab |
| Bouvante | 0-250 | 0.98 | 0.81 | 0.20 | 1.61 | 1.26 | Relab |
| Bouvante | 0-44 | 0.96 | 0.80 | 0.16 | 1.69 | 1.62 | Relab |
| Cachari | 0-25 | 0.96 | 0.86 | 0.43 | 1.97 | 6.40 | Relab |
| EET87520 | 0-45 | 1.01 | 0.88 | 0.49 | 2.19 | 3.91 | Relab |
| EET87542 | 0-25 | 0.98 | 0.89 | 0.30 | 1.28 | 4.11 | Relab |
| EET90020 | 0-25 | 0.98 | 0.88 | 0.52 | 1.97 | 6.83 | Relab |
| EET92003 | 0-125 | 0.93 | 0.90 | 0.31 | 2.20 | 4.04 | Relab |
| EETA79005 | 0-25 | 0.93 | 0.84 | 0.38 | 1.73 | 5.03 | Relab |
| EETA79005 | 0-250 | 0.94 | 0.89 | 0.26 | 1.81 | 2.82 | Relab |
| EETA79006 | 0-125 | 0.94 | 0.86 | 0.34 | 1.78 | 3.56 | Relab |
| GRO95533 | 0-25 | 0.97 | 0.87 | 0.51 | 1.81 | 3.38 | Relab |
| Jonzac | 0-25 | 0.94 | 0.81 | 0.38 | 1.99 | 4.25 | Relab |
| Juvinas | 0-25 | 0.95 | 0.86 | 0.42 | 1.77 | 3.08 | Relab |
| Juvinas | 125-250 | 0.98 | 0.99 | 0.13 | 2.02 | 1.15 | Relab |
| Juvinas | 25-45 | 0.95 | 0.91 | 0.27 | 2.19 | 2.99 | Relab |
| Juvinas | 45-75 | 0.96 | 0.95 | 0.20 | 2.26 | 2.44 | Relab |
| Juvinas | 75-125 | 0.97 | 0.97 | 0.15 | 2.17 | 1.71 | Relab |
| LEW85303 | 0-25 | 0.98 | 0.83 | 0.31 | 1.56 | 3.30 | Relab |
| LEW87004 | 0-25 | 0.94 | 0.84 | 0.35 | 1.63 | 4.37 | Relab |
| Macibini-clast3 | 0-63 | 0.96 | 0.87 | 0.21 | 1.71 | 1.61 | Relab |
| Millbillillie (C1MB69) | 0-25 | 0.98 | 0.93 | 0.29 | 1.95 | 1.32 | Relab |
| Millbillillie (C2MB69) | 0-25 | 0.98 | 0.93 | 0.29 | 2.01 | 1.53 | Relab |
| Millbillillie (CAMB69) | 0-25 | 0.96 | 0.87 | 0.43 | 1.72 | 1.97 | Relab |
| Millbillillie (C3MB69) | 0-25 | 0.97 | 0.85 | 0.43 | 1.74 | 2.12 | Relab |
| Millbillillie | 25-45 | 0.95 | 0.88 | 0.31 | 2.25 | 2.39 | Relab |
| Millbillillie | 45-75 | 0.96 | 0.89 | 0.25 | 2.26 | 1.92 | Relab |
| Millbillillie | 75-125 | 0.97 | 0.91 | 0.23 | 2.19 | 1.48 | Relab |
| Millbillillie | 0-80 | 0.97 | 0.89 | 0.38 | 1.44 | 1.86 | Relab |
| Millbillillie | 0-75 | 0.98 | 0.90 | 0.40 | 1.64 | 2.49 | Relab |
| Moore_County | 0-25 | 0.95 | 0.90 | 0.48 | 2.29 | 4.01 | Relab |
| PCA82501 | 0-125 | 0.98 | 0.86 | 0.26 | 1.42 | 2.40 | Relab |
| PCA82502 | 0-25 | 0.95 | 0.80 | 0.35 | 1.67 | 3.81 | Relab |
| PCA91006 | 0-125 | 0.98 | 0.91 | 0.35 | 1.90 | 3.77 | Relab |
| PCA91007 | 0-125 | 0.95 | 0.77 | 0.22 | 1.82 | 2.60 | Relab |
| PCA91078 | 0-45 | 0.99 | 0.83 | 0.28 | 1.80 | 2.78 | Relab |
| Padvarninkai | 0-25 | 0.95 | 0.84 | 0.35 | 1.73 | 2.48 | Relab |
| Padvarninkai | 25-45 | 0.96 | 0.91 | 0.26 | 1.95 | 1.93 | Relab |
| Pasamonte | 0-25 | 0.96 | 0.80 | 0.37 | 1.82 | 5.63 | Relab |
| Serra-de-Mage | 0-25 | 0.94 | 0.97 | 0.65 | 1.72 | 4.32 | Relab |
| Stannern | 0-25 | 0.95 | 0.83 | 0.35 | 1.55 | 2.80 | Relab |
| Stannern | 25-45 | 0.95 | 0.85 | 0.22 | 1.83 | 2.14 | Relab |
| Y-74450 | 0-25 | 0.95 | 0.87 | 0.32 | 1.54 | 3.01 | Relab |
| Y-74450 | 25-45 | 0.94 | 0.88 | 0.21 | 2.01 | 2.85 | Relab |
| Y-74450 | 45-75 | 0.95 | 0.92 | 0.18 | 2.18 | 2.14 | Relab |
| Y-74450 | 75-125 | 0.96 | 0.94 | 0.16 | 2.10 | 1.89 | Relab |
| Y-792510 | 0-25 | 0.97 | 0.85 | 0.43 | 1.87 | 3.59 | Relab |
| Y-792769 | 0-25 | 0.96 | 0.82 | 0.31 | 1.54 | 3.13 | Relab |
| Y-793591 | 0-25 | 0.96 | 0.82 | 0.37 | 1.55 | 3.80 | Relab |
| Y-82082 | 0-125 | 0.99 | 0.84 | 0.36 | 1.52 | 5.71 | Relab |
| **Dio** | **Size (μm)** | **BT** | **MR** | **R0.555** | **BS** | **VS** | **Source** |
| A-881526 | 0-25 | 0.82 | 1.01 | 0.54 | 2.65 | 10.98 | Relab |
| ALHA77256 | 0-25 | 0.84 | 0.95 | 0.48 | 2.23 | 8.48 | Relab |
| Aioun-el-Atrous | 0-25 | 0.86 | 0.97 | 0.62 | 2.54 | 6.42 | Relab |
| EETA79002 | 0-25 | 0.84 | 0.91 | 0.38 | 1.93 | 2.66 | Relab |
| EETA79002 | 125-250 | 0.87 | 1.01 | 0.22 | 2.28 | 1.30 | Relab |
| EETA79002 | 250-500 | 0.88 | 1.00 | 0.20 | 2.16 | 1.29 | Relab |
| EETA79002 | 25-45 | 0.81 | 0.96 | 0.28 | 2.73 | 2.36 | Relab |
| EETA79002 | 45-75 | 0.82 | 1.02 | 0.24 | 2.64 | 2.31 | Relab |
| EETA79002 | 75-125 | 0.85 | 1.06 | 0.22 | 2.55 | 1.75 | Relab |
| Ellemeet | 0-25 | 0.83 | 0.91 | 0.28 | 2.30 | 2.31 | Relab |
| GRO95555 | 0-25 | 0.84 | 0.95 | 0.61 | 2.51 | 10.17 | Relab |
| Johnstown | 0-25 | 0.87 | 0.94 | 0.40 | 1.75 | 4.23 | Relab |
| Johnstown | 25-45 | 0.83 | 1.07 | 0.35 | 2.57 | 4.55 | Relab |
| Johnstown | 0-75 | 0.83 | 1.06 | 0.50 | 2.70 | 7.27 | Relab |
| LAP91900 | 0-25 | 0.83 | 0.95 | 0.58 | 2.52 | 6.68 | Relab |
| Tatahouine | 0-25 | 0.82 | 0.97 | 0.61 | 2.61 | 5.33 | Relab |
| Y-74013 | 0-25 | 0.89 | 0.89 | 0.32 | 1.66 | 4.95 | Relab |
| Y-74013 | 25-45 | 0.85 | 0.93 | 0.29 | 2.42 | 4.62 | Relab |
| Y-75032 | 0-25 | 0.89 | 0.87 | 0.33 | 1.93 | 2.41 | Relab |
| Y-75032 | 25-45 | 0.86 | 0.92 | 0.20 | 2.70 | 2.82 | Relab |
| **How** | **Size (μm)** | **BT** | **MR** | **R0.555** | **BS** | **VS** | **Source** |
| Binda | 0-25 | 0.87 | 0.95 | 0.51 | 2.63 | 5.85 | Relab |
| Bununu | 0-25 | 0.93 | 0.88 | 0.41 | 1.72 | 2.52 | Relab |
| EET83376 | 0-25 | 0.93 | 0.84 | 0.42 | 1.71 | 4.87 | Relab |
| EET87503 (wet-sieved) | 0-150 | 0.93 | 0.88 | 0.26 | 1.65 | 1.91 | Relab |
| EET87503 | 0-25 | 0.93 | 0.88 | 0.39 | 1.56 | 2.32 | Relab |
| EET87503 (wet-sieved) | 0-25 | 0.94 | 0.87 | 0.32 | 1.43 | 2.75 | Relab |
| EET87503 | 125-250 | 0.96 | 0.96 | 0.19 | 1.73 | 0.91 | Relab |
| EET87503 | 250-500 | 0.97 | 0.95 | 0.18 | 1.64 | 0.75 | Relab |
| EET87503 | 25-45 | 0.93 | 0.89 | 0.30 | 1.75 | 1.90 | Relab |
| EET87503 | 45-75 | 0.93 | 0.94 | 0.23 | 1.90 | 1.14 | Relab |
| EET87503 | 75-125 | 0.95 | 0.97 | 0.20 | 1.87 | 1.11 | Relab |
| EET87513 | 0-25 | 0.94 | 0.87 | 0.34 | 1.50 | 3.10 | Relab |
| Frankfort | 0-25 | 0.90 | 0.85 | 0.37 | 1.95 | 2.01 | Relab |
| GRO95535 | 0-25 | 0.92 | 0.87 | 0.38 | 1.65 | 3.06 | Relab |
| GRO95574 | 0-125 | 0.92 | 0.90 | 0.29 | 1.83 | 1.81 | Relab |
| Kapoeta | 0-25 | 0.93 | 0.88 | 0.32 | 1.60 | 2.34 | Relab |
| Le-Teilleul | 0-25 | 0.89 | 0.87 | 0.40 | 2.09 | 3.35 | Relab |
| Petersburg | 0-25 | 0.94 | 0.86 | 0.32 | 1.67 | 9.11 | Relab |
| QUE94200 | 0-25 | 0.88 | 0.89 | 0.43 | 1.82 | 3.20 | Relab |

| Sample | Size (μm) | BT | MR | R0.555 | BS | VS | Source |
|---|---|---|---|---|---|---|---|
| QUE97001 | 0-150 | 0.87 | 0.95 | 0.31 | 2.16 | 3.00 | Relab |
| Y-7308 | 0-25 | 0.89 | 0.90 | 0.44 | 2.22 | 4.56 | Relab |
| Y-790727 | 0-25 | 0.91 | 0.86 | 0.36 | 1.80 | 6.25 | Relab |
| Y-791573 | 0-25 | 0.90 | 0.87 | 0.37 | 1.78 | 6.49 | Relab |
| NWA1949 | 0-63 | 0.95 | 0.79 | 0.29 | 1.58 | 4.11 | **This work** |
| DaG779 | 0-63 | 0.90 | 0.90 | 0.38 | 1.91 | 3.26 | **This work** |
| **CM-HED** | **Size (μm)** | **BT** | **MR** | **R0.555** | **BS** | **VS** | **Source** |
| NWA1949-10-Jbilet90 | 0-63 | 0.99 | 1.00 | 0.04 | 1.00 | 0.44 | **This work** |
| NWA1949-20-Jbilet80 | 0-63 | 0.99 | 0.99 | 0.05 | 1.01 | 0.46 | **This work** |
| NWA1949-30-Jbilet70 | 0-63 | 0.99 | 0.98 | 0.05 | 1.02 | 0.51 | **This work** |
| NWA1949-40-Jbilet60 | 0-63 | 0.99 | 0.97 | 0.05 | 1.03 | 0.57 | **This work** |
| NWA1949-50-Jbilet50 | 0-63 | 0.98 | 0.95 | 0.06 | 1.08 | 0.64 | **This work** |
| NWA1949-60-Jbilet40 | 0-63 | 0.98 | 0.94 | 0.07 | 1.10 | 0.79 | **This work** |
| NWA1949-70-Jbilet30 | 0-63 | 0.98 | 0.92 | 0.09 | 1.13 | 0.99 | **This work** |
| NWA1949-80-Jbilet20 | 0-63 | 0.97 | 0.90 | 0.12 | 1.18 | 1.31 | **This work** |
| NWA1949-90-Jbilet10 | 0-63 | 0.97 | 0.86 | 0.15 | 1.30 | 2.01 | **This work** |
| DAG779-10-Jbilet90 | 0-63 | 0.99 | 1.00 | 0.05 | 1.01 | 0.41 | **This work** |
| DAG779-20-Jbilet80 | 0-63 | 0.98 | 0.99 | 0.06 | 1.03 | 0.47 | **This work** |
| DAG779-30-Jbilet70 | 0-63 | 0.98 | 0.98 | 0.06 | 1.04 | 0.52 | **This work** |
| DAG779-40-Jbilet60 | 0-63 | 0.98 | 0.98 | 0.07 | 1.07 | 0.58 | **This work** |
| DAG779-50-Jbilet50 | 0-63 | 0.97 | 0.97 | 0.08 | 1.11 | 0.62 | **This work** |
| DAG779-60-Jbilet40 | 0-63 | 0.96 | 0.96 | 0.09 | 1.14 | 0.71 | **This work** |
| DAG779-70-Jbilet30 | 0-63 | 0.95 | 0.94 | 0.12 | 1.23 | 0.84 | **This work** |
| DAG779-80-Jbilet20 | 0-63 | 0.94 | 0.93 | 0.14 | 1.32 | 1.01 | **This work** |
| DAG779-90-Jbilet10 | 0-63 | 0.92 | 0.91 | 0.20 | 1.52 | 1.26 | **This work** |
| Millbillillie95-Murchison5 | 0-45 | 1.00 | 0.90 | 0.31 | 1.63 | 1.98 | HOSERLab |
| Millbillillie90-Murchison10 | 0-45 | 1.00 | 0.91 | 0.27 | 1.50 | 1.12 | HOSERLab |
| Millbillillie80-Murchison20 | 0-45 | 1.00 | 0.92 | 0.19 | 1.36 | 0.89 | HOSERLab |
| Millbillillie70-Murchison30 | 0-45 | 1.00 | 0.93 | 0.15 | 1.26 | 0.73 | HOSERLab |
| Millbillillie40-Murchison60 | 0-45 | 1.00 | 0.94 | 0.12 | 1.20 | 0.63 | HOSERLab |
| Millbillillie50-Murchison50 | 0-45 | 0.99 | 0.95 | 0.10 | 1.14 | 0.55 | HOSERLab |
| Millbillillie40-Murchison60 | 0-45 | 0.99 | 0.96 | 0.08 | 1.10 | 0.50 | HOSERLab |
| PRA04401 | 0-45 | 0.99 | 0.96 | 0.09 | 1.17 | 0.62 | HOSERLab |
| PRA04401 | 45-90 | 1.00 | 1.00 | 0.07 | 1.42 | 0.41 | HOSERLab |
| PRA04401 | 90-125 | 1.00 | 1.02 | 0.07 | 1.30 | 0.51 | HOSERLab |
| **Olivine-HED** | **Size (μm)** | **BT** | **MR** | **R0.555** | **BS** | **VS** | **Source** |
| NWA1949-10-Olivine90 | 0-63 | 1.06 | 0.97 | 0.65 | 1.36 | 3.33 | **This work** |
| NWA1949-20-Olivine80 | 0-63 | 1.03 | 0.93 | 0.56 | 1.37 | 3.06 | **This work** |
| NWA1949-30-Olivine70 | 0-63 | 1.02 | 0.90 | 0.49 | 1.38 | 3.52 | **This work** |
| NWA1949-40-Olivine60 | 0-63 | 1.01 | 0.89 | 0.46 | 1.38 | 3.72 | **This work** |
| NWA1949-50-Olivine50 | 0-63 | 1.00 | 0.87 | 0.42 | 1.41 | 3.72 | **This work** |
| NWA1949-60-Olivine40 | 0-63 | 0.98 | 0.85 | 0.39 | 1.43 | 3.95 | **This work** |
| NWA1949-70-Olivine30 | 0-63 | 0.98 | 0.84 | 0.35 | 1.48 | 3.86 | **This work** |
| NWA1949-80-Olivine20 | 0-63 | 0.97 | 0.82 | 0.32 | 1.52 | 4.07 | **This work** |
| NWA1949-90-Olivine10 | 0-63 | 0.96 | 0.82 | 0.32 | 1.53 | 3.92 | **This work** |
| DAG779-10-Olivine90 | 0-63 | 1.04 | 1.00 | 0.69 | 1.48 | 2.88 | **This work** |
| DAG779-20-Olivine80 | 0-63 | 1.02 | 0.98 | 0.62 | 1.49 | 2.91 | **This work** |
| DAG779-30-Olivine70 | 0-63 | 0.99 | 0.97 | 0.55 | 1.54 | 2.91 | **This work** |
| DAG779-40-Olivine60 | 0-63 | 0.97 | 0.96 | 0.49 | 1.57 | 2.73 | **This work** |
| DAG779-50-Olivine50 | 0-63 | 0.96 | 0.95 | 0.48 | 1.63 | 2.93 | **This work** |

| | | | | | | |
|---|---|---|---|---|---|---|
| DAG779-60-Olivine40 | 0-63 | 0.94 | 0.93 | 0.46 | 1.65 | 3.05 | **This work** |
| DAG779-70-Olivine30 | 0-63 | 0.92 | 0.93 | 0.43 | 1.78 | 3.03 | **This work** |
| DAG779-80-Olivine20 | 0-63 | 0.92 | 0.92 | 0.42 | 1.84 | 3.04 | **This work** |
| DAG779-90-Olivine10 | 0-63 | 0.90 | 0.91 | 0.38 | 1.91 | 2.81 | **This work** |
| NWA5480 | | 0.91 | 1.05 | 0.35 | 2.30 | 7.58 | Beck et al. 2011 |
| NWA4223 | | 0.95 | 0.99 | 0.16 | 1.39 | 2.96 | Beck et al. 2011 |
| NWA6013 | 0-45 | 0.95 | 0.92 | 0.31 | 1.69 | 4.00 | HOSERLab |
| NWA2968 | 0-45 | 1.02 | 0.97 | 0.16 | 1.06 | 3.66 | HOSERLab |
| **Olivine** | **Size (µm)** | **BT** | **MR** | **R0.555** | **BS** | **VS** | **Source** |
| Fo11 | 0-60 | 1.10 | 0.93 | 0.27 | 1.85 | 7.20 | USGS |
| Fo18 | 0-60 | 1.07 | 1.03 | 0.35 | 1.76 | 8.01 | USGS |
| Fo29 | 0-60 | 1.16 | 1.03 | 0.33 | 2.34 | 9.02 | USGS |
| Fo41 | 0-60 | 1.10 | 0.99 | 0.36 | 1.80 | 7.75 | USGS |
| Fo51 | 0-60 | 1.12 | 1.00 | 0.38 | 1.85 | 6.60 | USGS |
| Fo60 | 0-60 | 1.14 | 1.01 | 0.46 | 1.98 | 7.83 | USGS |
| Fo66 | 0-60 | 1.17 | 1.01 | 0.44 | 2.28 | 7.93 | USGS |
| Fo89 | 0-60 | 1.10 | 0.96 | 0.58 | 1.40 | 5.94 | USGS |
| Fo91 | 0-60 | 1.26 | 1.01 | 0.55 | 2.50 | 8.36 | USGS |
| Fo10 | 0-45 | 1.03 | 1.03 | 0.58 | 1.30 | 4.56 | USGS |
| Fo20 | 0-45 | 1.08 | 1.06 | 0.68 | 1.57 | 4.03 | USGS |
| Fo30 | 0-45 | 1.11 | 1.08 | 0.70 | 1.81 | 6.25 | USGS |
| Fo40 | 0-45 | 1.07 | 0.96 | 0.36 | 1.24 | 5.76 | USGS |
| Fo50 | 0-45 | 1.11 | 1.05 | 0.57 | 1.85 | 7.40 | USGS |
| Fo60 | 0-45 | 1.12 | 1.02 | 0.49 | 1.92 | 8.85 | USGS |
| Fo70 | 0-45 | 1.06 | 0.90 | 0.17 | 1.13 | 5.38 | USGS |
| Fo80 | 0-45 | 1.08 | 0.96 | 0.27 | 1.34 | 5.53 | USGS |
| Fo90 | 0-45 | 1.07 | 0.96 | 0.19 | 1.26 | 4.20 | USGS |
| Fo91 | 0-38 | 1.07 | 1.01 | 0.60 | 1.27 | 7.54 | HOSERLab |
| Fo90 | 0-63 | 1.11 | 1.04 | 0.88 | 1.57 | 9.52 | **This work** |
| **CM** | **Size (µm)** | **BT** | **MR** | **R0.555** | **BS** | **VS** | **Source** |
| ALH84044 | 0-125 | 0.98 | 0.95 | 0.05 | 0.99 | 0.87 | Relab |
| LEW87148 | 0-125 | 0.99 | 0.96 | 0.05 | 1.04 | 0.69 | Relab |
| GRO85202 | 0-125 | 0.99 | 0.97 | 0.05 | 1.04 | 0.89 | Relab |
| ALHA77306 | 0-125 | 0.98 | 0.97 | 0.05 | 1.05 | 1.03 | Relab |
| LEW87022 | 0-125 | 0.99 | 0.96 | 0.04 | 0.98 | 0.56 | Relab |
| EET87522 | 0-125 | 0.99 | 1.00 | 0.03 | 0.98 | 0.27 | Relab |
| MAC88100 | 0-125 | 1.00 | 1.00 | 0.03 | 1.03 | 0.30 | Relab |
| LON94101 | 0-125 | 0.99 | 0.98 | 0.04 | 1.03 | 0.58 | Relab |
| Y-791191 | 0-125 | 0.99 | 0.97 | 0.05 | 1.00 | 0.47 | Relab |
| Y-791824 | 0-125 | 0.99 | 0.97 | 0.04 | 1.07 | 0.68 | Relab |
| A-881458 | 0-125 | 0.99 | 0.99 | 0.05 | 1.00 | 0.48 | Relab |
| A-881955 | 0-125 | 0.99 | 1.00 | 0.05 | 1.04 | 0.79 | Relab |
| A-881280 | 0-125 | 1.01 | 0.97 | 0.05 | 0.99 | 0.61 | Relab |
| A-881594 | 0-125 | 1.00 | 0.99 | 0.04 | 1.00 | 0.16 | Relab |
| Y-82054 | 0-125 | 1.01 | 1.01 | 0.04 | 1.00 | 0.41 | Relab |
| MET00639 | 0-75 | 1.00 | 0.98 | 0.03 | 0.99 | 0.03 | Relab |
| Cold-Bokkeveld | 0-75 | 0.97 | 1.00 | 0.05 | 1.01 | 0.33 | Relab |
| Mighei | 0-75 | 0.98 | 0.99 | 0.05 | 0.99 | 0.60 | Relab |
| Jbilet | 0-63 | 0.99 | 1.00 | 0.04 | 0.99 | 0.39 | **This work** |
| Murchison | 0-40 | 0.98 | 0.96 | 0.04 | 0.99 | 0.64 | HOSERLab |
| Murchison | 0-45 | 0.99 | 0.99 | 0.05 | 0.98 | 0.37 | HOSERLab |
| A-881955 | 0-125 | 0.99 | 1.00 | 0.05 | 1.04 | 0.78 | HOSERLab |
| LEW87016 | 0-125 | 0.98 | 1.04 | 0.05 | 1.07 | 0.54 | HOSERLab |
| Y791191 | 0-125 | 1.00 | 0.97 | 0.05 | 1.00 | 0.45 | HOSERLab |
| Y74662 | 0-125 | 1.00 | 1.00 | 0.05 | 1.03 | 0.31 | HOSERLab |
| Mighei-bulk | 0-40 | 0.99 | 0.99 | 0.04 | 0.99 | 0.37 | HOSERLab |
| **Olivine-(rich)-Opx** | **Size (µm)** | **BT** | **MR** | **R0.555** | **BS** | **VS** | **Source** |

| mixture | | | | | | | |
|---|---|---|---|---|---|---|---|
| Olivine50-Opx50 | 0-38 | 0.93 | 1.09 | 0.56 | 1.85 | 4.33 | HOSERLab |
| Olivine50-Opx50 | 38-53 | 0.94 | 1.15 | 0.43 | 2.48 | 8.45 | HOSERLab |
| Olivine50-Opx50 | 63-90 | 0.93 | 1.29 | 0.38 | 3.11 | 4.42 | HOSERLab |
| Olivine50-Opx50 | 90-125 | 0.96 | 1.26 | 0.27 | 2.85 | 3.09 | HOSERLab |
| Olivine60-Opx40 | 0-38 | 0.96 | 1.07 | 0.57 | 1.72 | 4.20 | HOSERLab |
| Olivine60-Opx40 | 38-53 | 0.93 | 1.15 | 0.44 | 2.50 | 8.40 | HOSERLab |
| Olivine60-Opx40 | 63-90 | 0.95 | 1.27 | 0.37 | 2.83 | 8.46 | HOSERLab |
| Olivine60-Opx40 | 90-125 | 0.97 | 1.28 | 0.26 | 2.89 | 6.19 | HOSERLab |
| Olivine70-Opx30 | 0-38 | 0.99 | 1.05 | 0.59 | 1.55 | 4.09 | HOSERLab |
| Olivine70-Opx30 | 38-53 | 0.98 | 1.12 | 0.43 | 2.26 | 8.69 | HOSERLab |
| Olivine70-Opx30 | 63-90 | 1.04 | 1.19 | 0.39 | 2.54 | 9.75 | HOSERLab |
| Olivine70-Opx30 | 90-125 | 1.05 | 1.16 | 0.27 | 2.36 | 6.60 | HOSERLab |
| Olivine80-Opx20 | 0-38 | 1.02 | 1.03 | 0.59 | 1.41 | 3.95 | HOSERLab |
| Olivine80-Opx20 | 38-53 | 1.03 | 1.09 | 0.42 | 2.06 | 8.57 | HOSERLab |
| Olivine80-Opx20 | 63-90 | 1.06 | 1.16 | 0.40 | 2.55 | 10.51 | HOSERLab |
| Olivine80-Opx20 | 90-125 | 1.10 | 1.10 | 0.27 | 2.20 | 7.03 | HOSERLab |
| Olivine90-Opx10 | 0-38 | 1.04 | 1.02 | 0.59 | 1.34 | 7.52 | HOSERLab |
| Olivine90-Opx10 | 38-53 | 1.10 | 1.05 | 0.41 | 1.96 | 8.82 | HOSERLab |
| Olivine90-Opx10 | 63-90 | 1.13 | 1.10 | 0.42 | 2.39 | 11.34 | HOSERLab |
| Olivine90-Opx10 | 90-125 | 1.13 | 1.08 | 0.25 | 2.16 | 6.48 | HOSERLab |
| **Melt/glass/shock** | **Size (μm)** | **BT** | **MR** | **R0.555** | **BS** | **VS** | **Source** |
| Padvarninkai (+melt) | 0-25 | 0.99 | 0.89 | 0.17 | 1.22 | 1.25 | Relab |
| Padvarninkai (+melt) | 25-45 | 1.00 | 0.91 | 0.14 | 1.19 | 0.81 | Relab |
| LEW85303 (+melt) | 0-25 | 0.98 | 0.83 | 0.31 | 1.56 | 3.30 | Relab |
| Macibini-cl.3-melt (glass) | | 1.06 | 1.05 | 0.37 | 2.50 | 12.42 | Relab |
| JaH626-shock | 250-500 | 1.08 | 0.87 | 0.12 | 1.56 | 0.52 | HOSERLab |
| JaH626-shock | 45-90 | 1.11 | 0.78 | 0.15 | 1.81 | 1.42 | HOSERLab |
| JaH626-shock | 90-125 | 1.10 | 0.83 | 0.11 | 1.79 | 0.71 | HOSERLab |
| JaH626-shock | <250 | 1.12 | 0.77 | 0.20 | 1.60 | 1.77 | HOSERLab |
| JaH626-shock | <45 | 1.11 | 0.77 | 0.28 | 1.63 | 3.06 | HOSERLab |

**Appendix 4: Finding FC data points that belong to a particular polyhedron created using band parameter values of laboratory spectra:**

We use an expanding volume approach i.e., an algorithm in IDL programming language to test whether a data point does or does not belong to that particular polyhedron. The volume of the polyhedron is computed using an array of vertices (the band parameter values) and a connectivity array defined through the convex hull method. For example, Volume$_{lab\_polyhedron}$ represents the volume using the arrays of Vertices$_{lab\_polyhedron}$ and Connectivity$_{lab\_polyhedron}$ i.e.,

$$\text{Vertices}_{lab\_polyhedron} = \begin{bmatrix} x_1 & y_1 & z_1 \\ x_2 & y_2 & z_2 \\ \ldots & \ldots & \ldots \\ \ldots & \ldots & \ldots \\ \ldots & \ldots & \ldots \\ x_n & x_n & x_n \end{bmatrix}, \text{Connectivity}_{lab\_polyhedron} = \begin{bmatrix} a_1 & b_1 & c_1 & d_1 \\ a_2 & b_2 & c_2 & d_2 \\ \ldots & \ldots & \ldots & \ldots \\ \ldots & \ldots & \ldots & \ldots \\ \ldots & \ldots & \ldots & \ldots \\ a_n & b_n & c_n & d_n \end{bmatrix}$$

New polyhedrons are created with each and every pixel of the FC data appending the band parameter values to the existing array of vertices. The volume of the new polyhedron is calculated using the new array of vertices and the new connectivity array. Suppose that, for a particular FC data point, Volume$_{new\_polyhedron}$ represents the new volume using Vertices$_{new\_polyhedron}$ and Connectivity$_{new\_polyhedron}$ i.e.,

$$\text{Vertices}_{new\_polyhedron} = \begin{bmatrix} x_1 & y_1 & z_1 \\ x_2 & y_2 & z_2 \\ \ldots & \ldots & \ldots \\ \ldots & \ldots & \ldots \\ \ldots & \ldots & \ldots \\ x_n & x_n & x_n \\ x_{FC} & y_{FC} & z_{FC} \end{bmatrix}, \text{Connectivity}_{new\_polyhedron} = \begin{bmatrix} p_1 & q_1 & r_1 & s_1 \\ p_2 & q_2 & r_2 & s_2 \\ \ldots & \ldots & \ldots & \ldots \\ \ldots & \ldots & \ldots & \ldots \\ \ldots & \ldots & \ldots & \ldots \\ p_n & q_n & r_n & s_n \end{bmatrix}$$

For each and every pixel in the FC dataset, the volumes are compared. If the new volume (of the new polyhedron) is less than the volume of the laboratory polyhedron, i.e.,

Volume$_{new\_polyhedron}$ > Volume$_{lab\_polyhedron}$, then the FC data point is inside the laboratory polyhedron.

Table 1: Band parameters used in this study.

| Band Parameters | Definition |
|---|---|
| Band Tilt (BT) | $R_{0.92}/R_{0.96}$ |
| Mid Ratio (MR) | $(R_{0.75}/R_{0.83})/(R_{0.83}/R_{0.92})$ |
| Band Strength (BS) | $R_{max\_VIS}/R_{min\_NIR}$ |
| Visible Slope (VS) in % per 0.1μm | $(R_{\lambda 2}-R_{\lambda 1})*10/(\lambda_2-\lambda_1)$ |

where, R is the reflectance value at the given wavelengths in μm; $R_{max\_VIS}$ and $R_{min\_NIR}$ are the maximum reflectance in the visible wavelength range and the minimum reflectance in the near-infrared wavelength range, respectively.

**Figure captions**

Fig. 1. (A) Absolute and (B) normalized laboratory spectra (line) resampled to FC bandpasses (symbols). The spectra of NWA 1929 howardite (eucrite-rich), DaG 779 howardite (diogenite-rich), Jbilet-Winselwan (CM2-chondrite), and olivine are measured in this work. The Macibini-eucrite glass spectrum is from RELAB. (C) Sketch illustrating the band parameters calculated from the absolute reflectance spectrum, Band Tilt (BT), Mid Ratio (MR), Visible Slope in % per 0.1 μm (VS) and Band Strength (BS). The acronyms $R_{min\_VIS}$ and $R_{max\_VIS}$ represent those FC color filters showing minimum and maximum reflectance in the visible wavelength range, while $R_{min\_NIR}$ is the minimum reflectance in the near-infrared wavelength range.

Fig. 2. Resampled spectra of Jbilet-Winselwan (CM2-chondrite) mixed with howardites DaG 779 and NWA 1929. (A, B) Absolute reflectance spectra in an interval of 10 wt.%, represented by the dotted lines. (C, D) Normalized spectra of samples and mixtures in steps of 10, 30, 50, 70, 90 wt.%.

Fig. 3. Resampled spectra of olivine mixed with howardites DaG 779 and NWA 1929. (A, B) Absolute reflectance spectra in an interval of 10 wt.%, represented by the dotted lines. (C, D) Normalized spectra of samples and mixtures in steps 10, 30, 50, 70, 90 wt.%.

Fig. 4: Two-dimensional band parameter spaces for DaG 779, NWA 1929, olivine, Jbilet-Winselwan, Macibini-eucrite glass, and mixtures of howardites with olivine or Jbilet-Winsewlan. (A) VS versus $R_{0.55\mu m}$, (B) BS versus $R_{0.55\mu m}$, (C, D) polygons defined by all the spectra of eucrites, diogenites, howardites, olivine, and CM2 in the corresponding band parameter spaces.

Fig. 5: Different perspective views of three-dimensional band parameter spaces. (A) BT versus MR versus $R_{0.55\mu m}$, (B) BT versus MR versus VS, (C) BT versus MR versus BS, (D) VS versus BS versus $R_{0.55\mu m}$. Each polyhedron represents a particular mineralogy (Euc: eucrite, Dio: diogenite, How: howardite, Ol: olivine, Ol-HED: olivine plus howardite mixtures, and olivine-bearing diogenites, CM2: CM2 chondrites, CM2-HED: Jbilet plus howardite mixture, Murchison plus eucrite mixture, and CM2-bearing howardite). The orange data point represents the Macibini eucrite glass sample.

Fig. 6: Arruntia region from FC HAMO data (~ 60 m/pix). (A) Reflectance image at 0.55 μm. Derived band parameters: (B) MR (Mid Ratio), (C) VS (Visible Slope in % per 0.1μm), (D) BS (1-μm band strength), (E) BT (Band Tilt), (F) Surface (topographic) slope in degrees computed from HAMO-DTM (~ 62 m/pixel).

Fig. 7: Mineralogy of Arruntia based on our laboratory polyhedrons of BT versus MR versus $R_{0.55\mu m}$: (A) Eucrites, (B) CM2 plus HED mixtures (≤ 20-30 wt.% of CM2, indistinguishable from eucrites/howardites), (C) Olivine plus HED mixtures (≤ 40-60 wt.% of olivine, indistinguishable from eucrites/howardites), (D) Olivine-rich regions (≥ 40-60 wt.% olivine, distinguished from eucrites/howardites).

Fig. 8: Potential impact-melt flow features (marked by arrows) observed in Arruntia and its ejecta. North is upwards for all figures. (A) HAMO image at 0.55 μm ( ~ 60 m/pixel) showing the locations of the features identified from LAMO images (~ 18 m/pixel). (B) dark streaks on the bright ejecta near the outer rim, (C, D) lobate flow features on the slopes of inner crater wall, (E) potential impact-melt deposits partly ponded, imparting a darker tone on the brighter debris in the floor of the crater (enhanced image showing a scene that is only illuminated by multi-scattered light), (F) lobate flow-features in the relatively brighter ejecta overlying the densely cratered older surface.

Fig. 1.

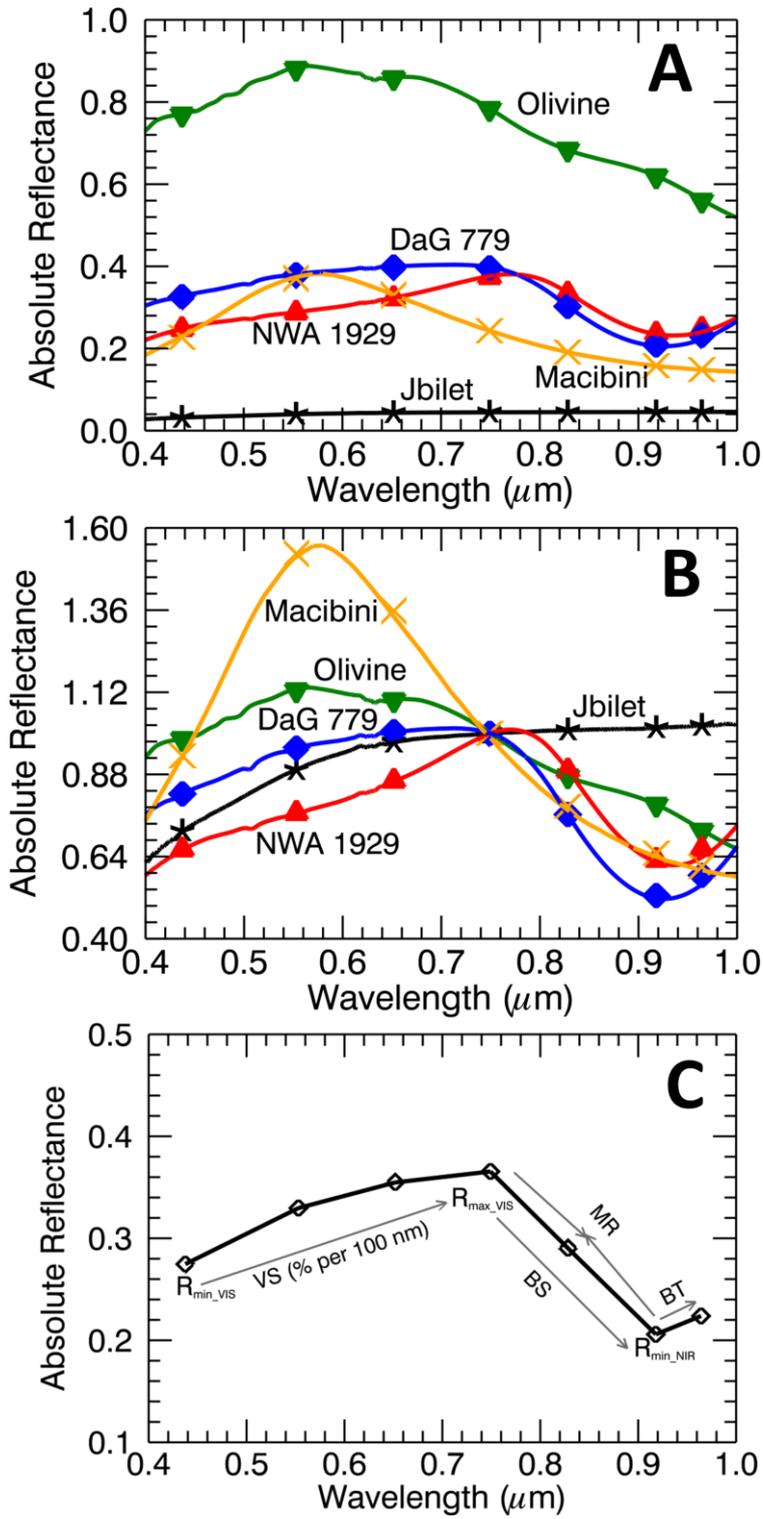

Fig. 2

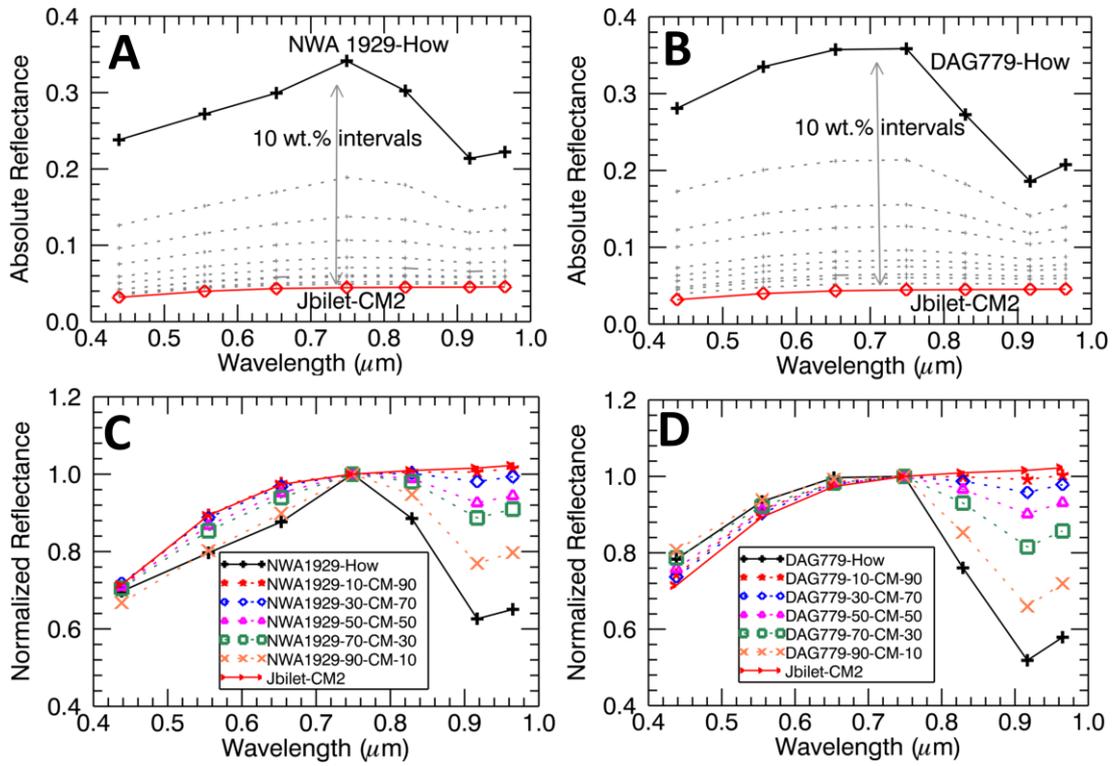

Fig. 3

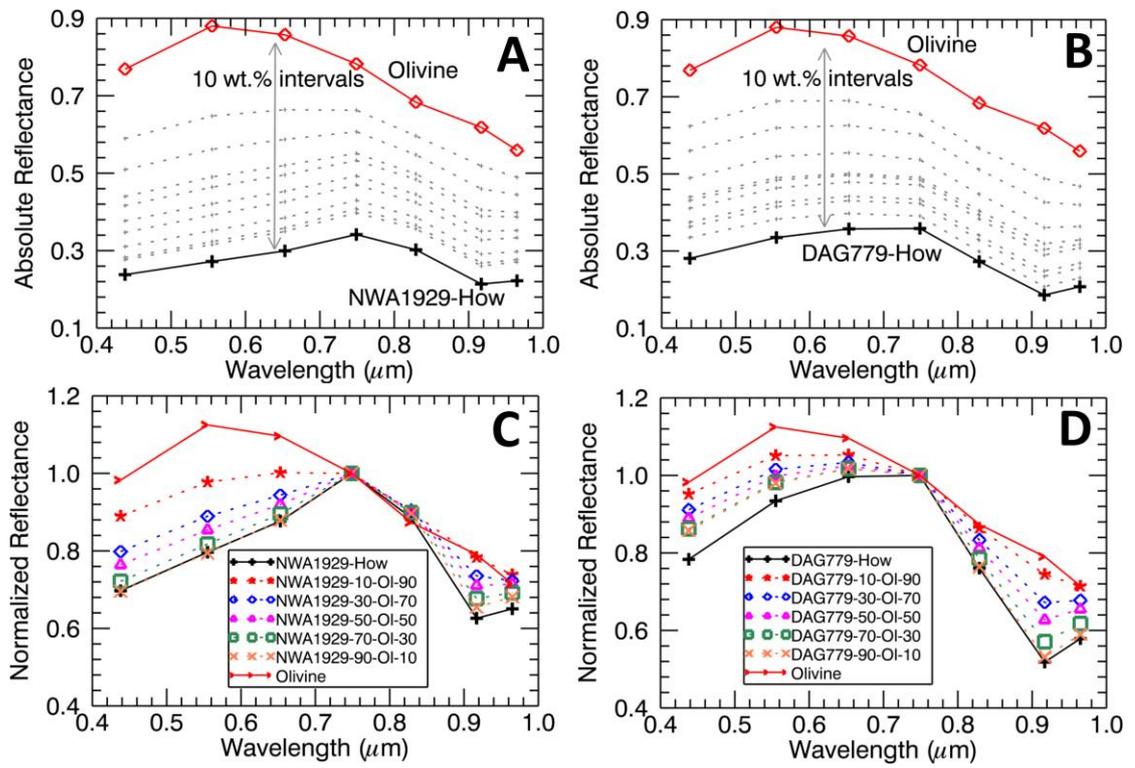

Fig. 4

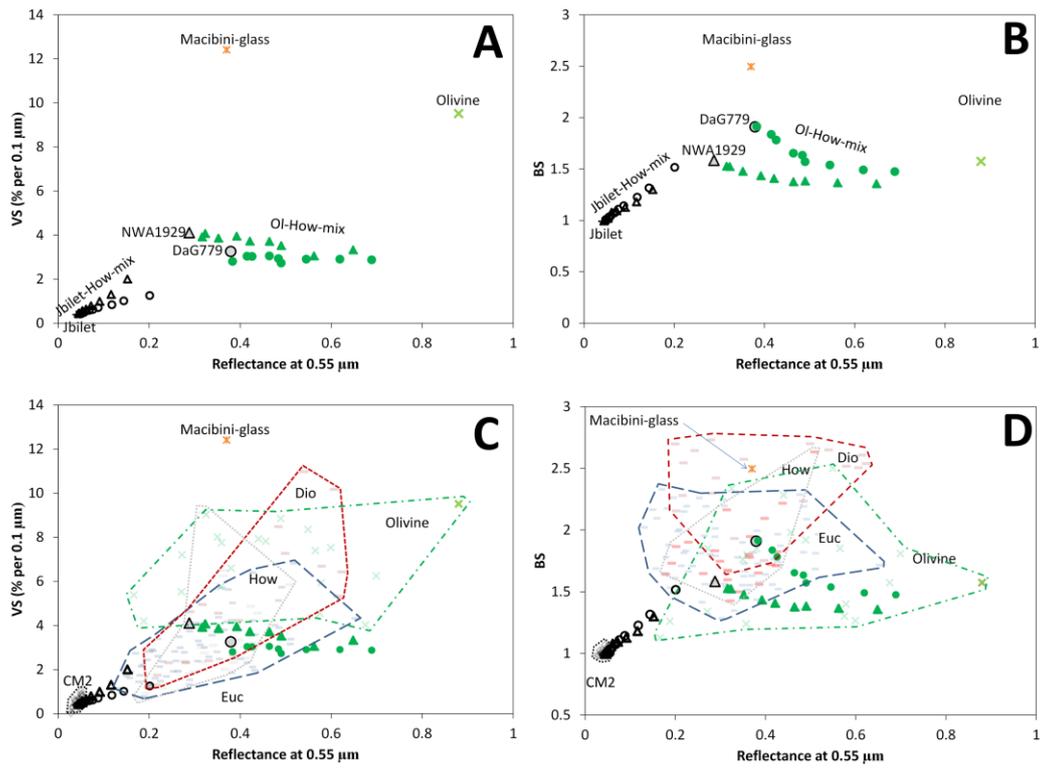

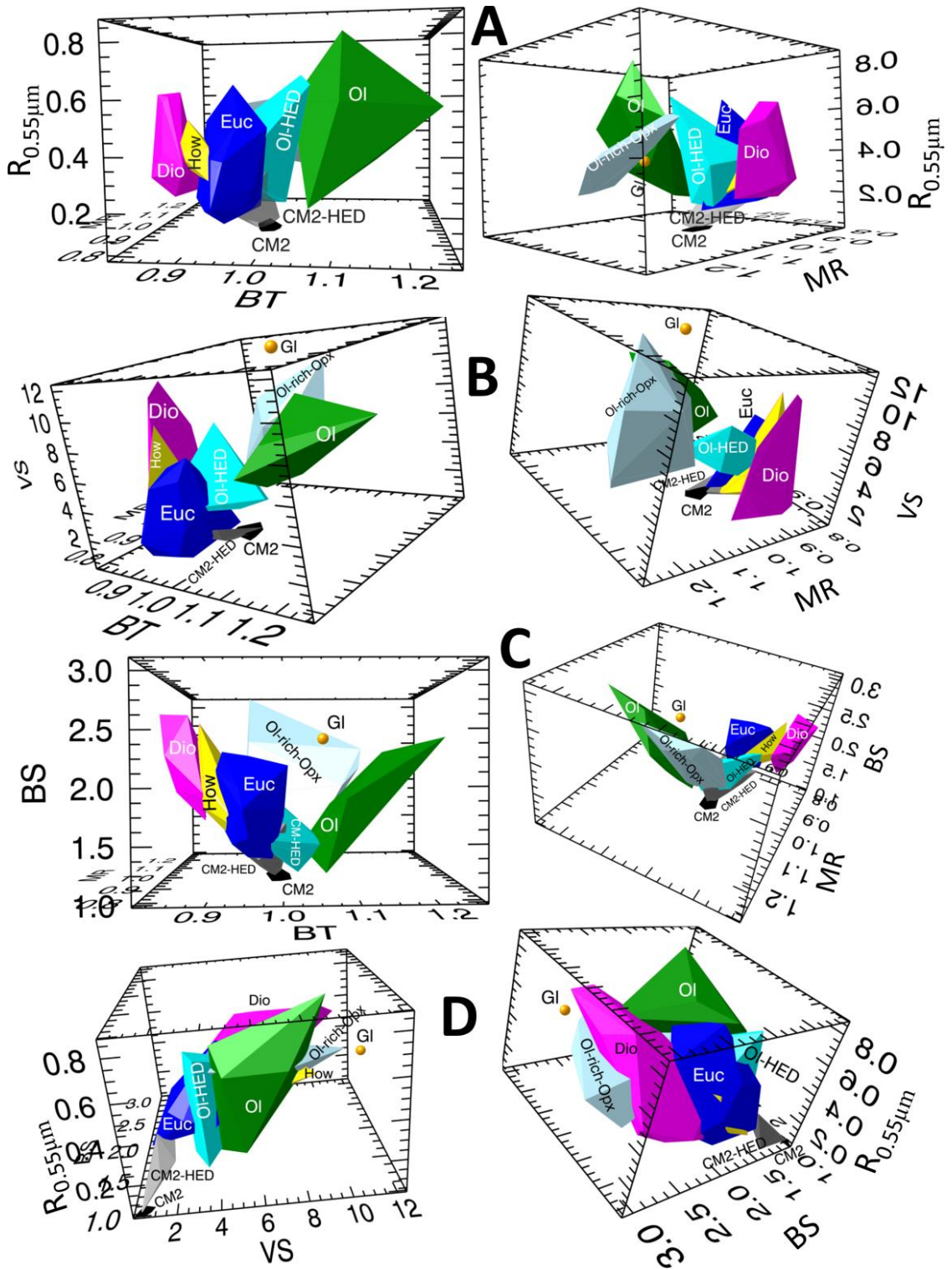

Fig. 6

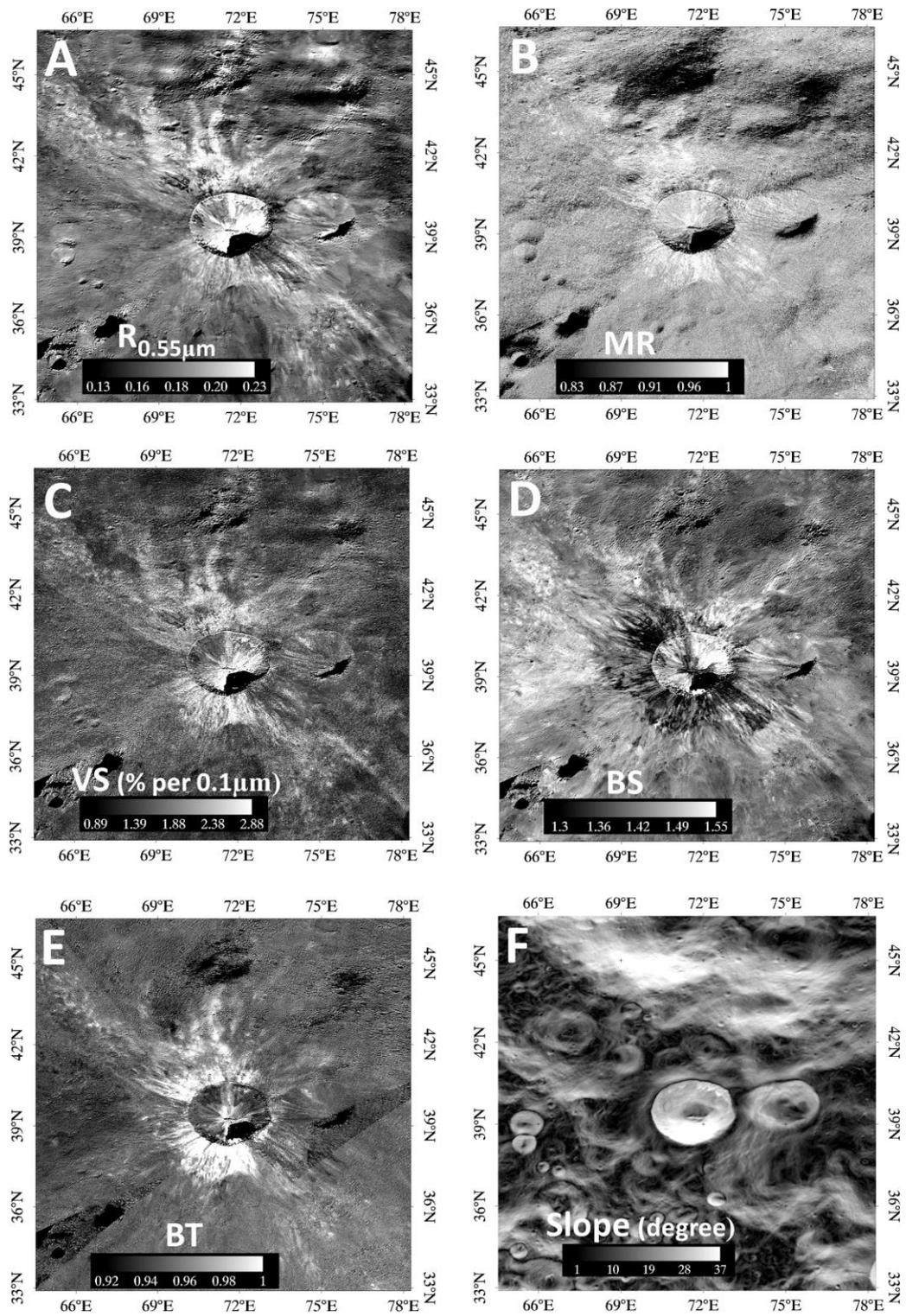

Fig. 7

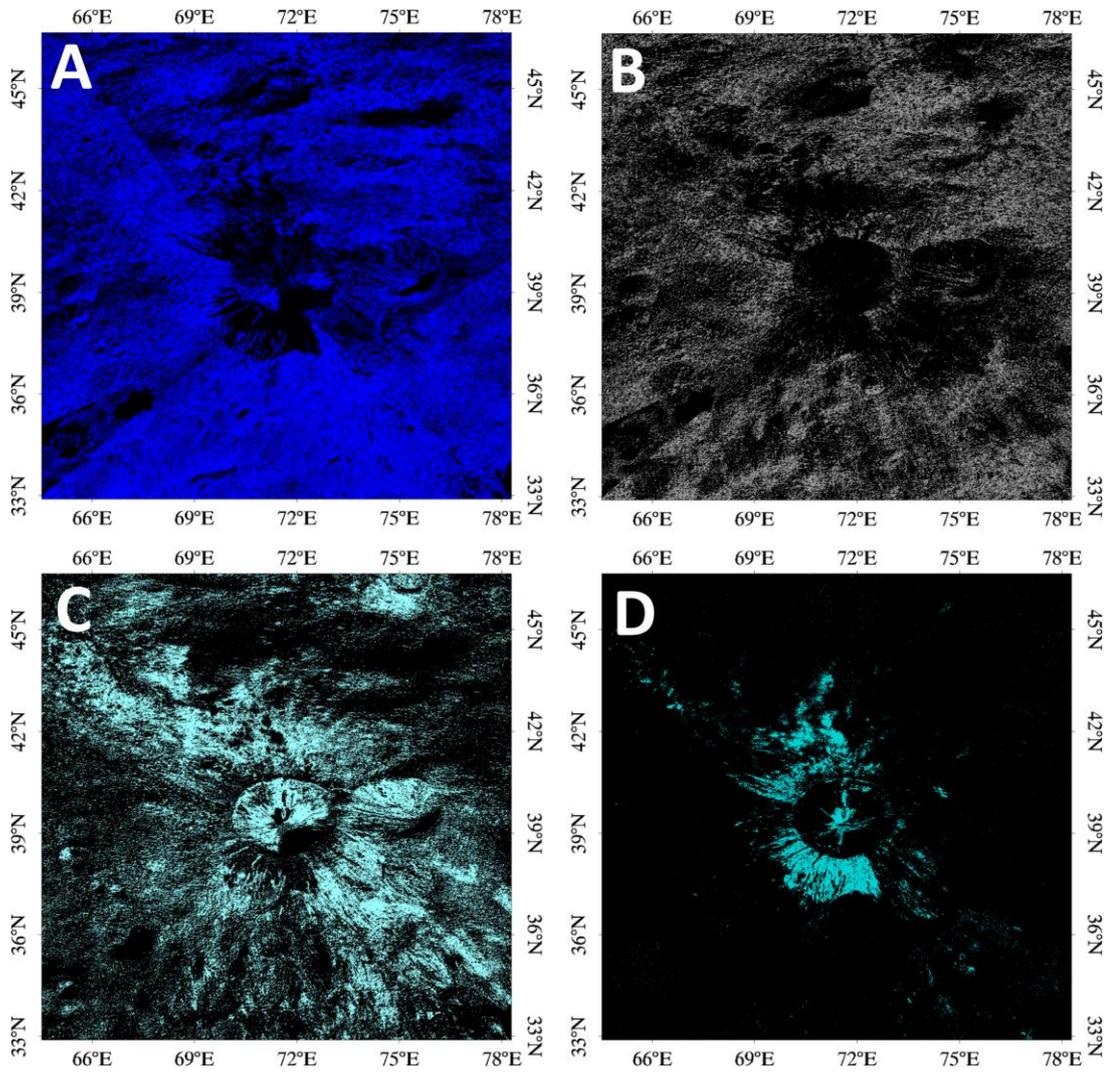

Fig. 8

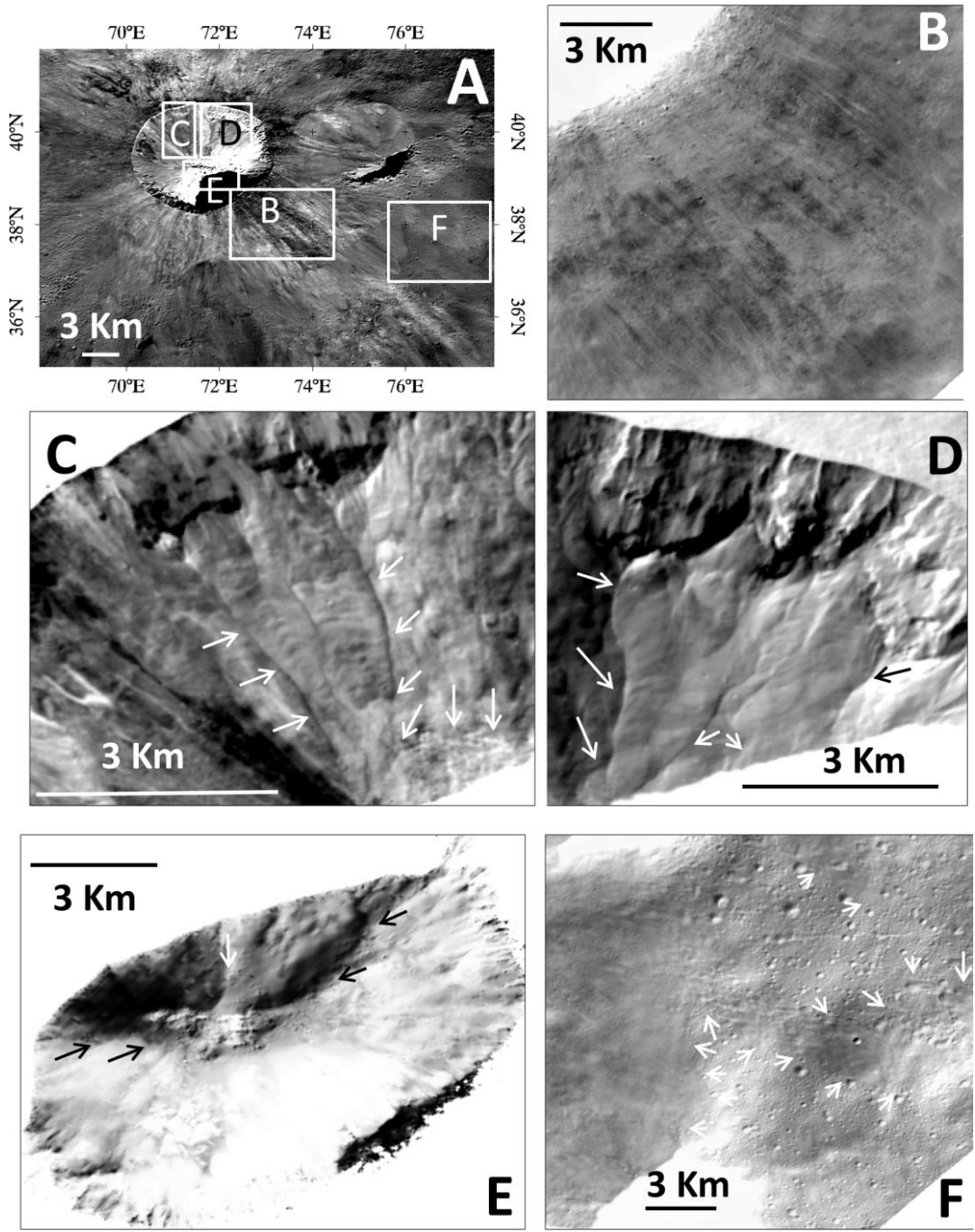